\documentclass{jfm}
\usepackage{graphicx}
\usepackage{epstopdf, epsfig}
\usepackage{xcolor}
\usepackage{psfrag}
\usepackage{tikz}
\usepackage{multirow}
\usepackage{amsmath}
\usepackage{amssymb}

\newcommand{\Me}{M\!e\,}

\definecolor{light-red}{rgb}{1,0.5,0.5}
\definecolor{light-blue}{rgb}{0.5,0.5,1}
\shorttitle{Surface waves of a walker}
\shortauthor{Lo\"ic Tadrist, Jeong-Bo Shim, Tristan Gilet and Peter Schlagheck}

\title{Faraday instability and subthreshold Faraday waves: surface waves emitted by walkers}

\author{Lo\" ic Tadrist\aff{1},
  Jeong-Bo Shim\aff{2},
  Tristan Gilet\aff{1}
 \and Peter Schlagheck\aff{2}\corresp{\email{Loic.tadrist@uliege.be}}}

\affiliation{\aff{1} Microfluidics Lab, Department of Mechanical and Aerospace Engineering, University of Liege,
All\' ee de la d\' ecouverte 9, 4000 Li\`ege, Belgium \aff{2}IPNAS, CESAM research unit, University of Liege, Allee du 6 Ao\^ut 15, 4000 Li\`ege, Belgium}

\begin{document}

\maketitle

\begin{abstract}
A walker is a fluid entity comprising a bouncing droplet coupled to the waves that it generates at the surface of a vibrated bath. Thanks to this coupling, walkers exhibit a series of wave-particle features formerly thought to be exclusive to the quantum realm. In this paper, we derive a model of the Faraday surface waves generated by an impact upon a vertically vibrated liquid surface. We then particularise this theoretical framework to the case of forcing slightly below the Faraday instability threshold. Among others, this theory yields a rationale for the dependence of the wave amplitude to the phase of impact, as well as the characteristic timescale and length scale of viscous damping. The theory is validated with experiments of bead impact on a vibrated bath. We finally discuss implications of these results for the analogy between walkers and quantum particles.
\end{abstract}

\begin{keywords}
Faraday instability, Faraday waves, Impact on free surface, Bouncing droplet, Walkers.
\end{keywords}

\section{Introduction}

\begin{table}
\begin{center}
\begin{tabular}{lllllllll}
\multicolumn{9}{l}{\textsc{General parameters}}\\ 
&&&&&&&&\\
3D vector position& \multicolumn{2}{l}{$\vec{r}= (x,y,z)$}&&& 2D vector position &\multicolumn{2}{l}{$\mathbf{r}=(x,y)$}&\\
Radial distance & $r$&&&& Time &$t$&&\\
Dimensionless time &\multicolumn{2}{l}{$\tau=\Omega t/2$}&&& Laplace variable& $s$&&\\
Wavelength& $\lambda$&&&& Wavenumber&$k$&&\\
%Bessel function & $J_0$&&&&&&&\\
&&&&&&&&\\
\multicolumn{9}{l}{\textsc{Fields}}\\ 
&&&&&&&&\\
Pressure field&$P(\vec{r},t)$&&&& Velocity field& $\vec{v}(\vec{r},t)$&&\\
Vertical velocity&$v_z(\vec{r},t)$&&&&Surface wave profile &$\zeta(\mathbf{r},t)$&&\\
&&&&&&&&\\
\multicolumn{9}{l}{\textsc{Fluid parameters}}\\ 
&&&&&&&&\\
Gravity acceleration &$g$& 9.81&m.s$^{—2}$&&Density&$\rho$&956&kg/m$^{3}$\\
Surface tension&$\sigma$&$20.6$ &mN/m&&Viscosity&$\nu$&$20$ &cS \\
&&&&&&&&\\
\multicolumn{9}{l}{\textsc{Input parameters}}\\ 
&&&&&&&&\\
Shaker pulsation & $\Omega$& $2\pi\times 80$& m$^{-1}$&&Driving strength & $\Gamma$& & \\
Walking velocity&$V_w$ & 11 &mm/s&&\textit{reduced form}, $v$&$2 V_w/\Omega$& 43.8& $\mu$m\\
Impact phase & $\tau_i$& & &&&&&\\
&&&&&&&&\\
\multicolumn{9}{l}{\textsc{Output parameters}}\\ 
&&&&&&&&\\
Faraday period& $T_F$& 25&ms&& Faraday threshold & $\Gamma_F$&4.155 &\\
Faraday wavelength&$\lambda_F$& $4.749$& mm&& Faraday wavenumber& $k_F$& 1323 & m$^{-1}$\\
Phase shift& $\theta_F$& $\pi/4$ & &&Damping length& $l$& $10.2$& mm \\
Damping time & $1/\delta_F$& -49&&&\textit{dimensional form}&$2/\Omega \delta_F$ & 195& ms\\
Diffusion coefficient& $D$& $2.20$& mm$^2$&&\textit{dimensional form}&$\Omega D/2$ & $5.5$ &cm$^2$/s\\
Group velocity& $c_F$& $0.983$& mm&&\textit{dimensional form}&$\Omega c_F/2$ &24.7& cm/s\\
&&&&&&&&\\
\multicolumn{9}{l}{\textsc{Dimensionless parameters}}\\ 
&&&&&&&&\\
Free pulsation &$\omega_0$& 1&&& Free damping& $\gamma_0$& 0.2771&\\
Threshold distance&$\mathcal{M}$ & 10 &&&Memory number&$\Me$& 7.65&\\
\end{tabular}
\end{center}
\caption{Table of notations used along the paper. Indicice $_0$ (respectively $_F$ and $_i$) refers  to the solution for a null viscosity (respectively to the solution at the Faraday threshold and to the quantities relative to the impact). }\label{tab:notations}
\end{table}

Michael \citet{Faraday1831} gave a first experimental description of the instability that would be named after him. The Faraday instability occurs when a container filled with liquid is vertically vibrated with an acceleration $\Gamma g \cos (\Omega t)$. When $\Gamma$ is higher than a threshold $\Gamma_F $ the forcing is sufficient to counterbalance viscous dissipation \citep{Kumar1994}, the initially flat free surface destabilises and beautiful patterns of standing waves spontaneously appear. The resonance is often subharmonic so the frequency of the standing waves is $\Omega/4\pi$. The Faraday wavelength $\lambda_F$, defined as the characteristic wavelength of the standing waves, is usually close to the wavelength prescribed by the dispersion relation of gravity-capillary waves with a frequency $\Omega/4\pi$. In its original and now most studied configuration \citep{Benjamin1954}, the depth and lateral extension of the liquid bath are respectively much deeper and much larger than the wavelength, which ensures that the resonance is subharmonic. 

Below the Faraday instability threshold, all surface waves are damped and, without any perturbation, the liquid surface remains flat. A perturbation can be added in the form of a bouncing droplet. If the vertical vibration of the bath is of sufficient amplitude, this droplet can bounce on the bath periodically and indefinitely \citep{Couder2005, Gilet2008, Molacek2013a}. \citet{Couder2005b} showed that such droplets may excite Faraday waves, especially if they can bounce once every two forcing periods (i.e. at frequency $\Omega/4\pi$). If the forcing amplitude is slightly below the instability threshold ($\Gamma \lesssim \Gamma_F$), these waves are slowly damped and they are still present when the droplet impacts again \citep{Eddi2011}. Consequently, the droplet impacts on a non-flat surface and some horizontal momentum can be transferred from the wave to the droplet. The latter starts to ``walk'' on the vibrated liquid surface. The combination of a walking droplet and the underlying surface waves is called a walker. A typical walker involves a droplet of radius $R_d = 0.4$~mm and a bath of depth 6~mm, both made of silicone oil of density $\rho = 956$kg~m$^{-3}$, kinematic viscosity $\nu = 20$~cS and surface tension $\sigma = 20.6$~mN~m$^{-1}$. The bath is typically shaken at a frequency $\Omega / (2\pi) = 80$~Hz with an amplitude $\Gamma \sim 4$ (table \ref{tab:notations}).  

Walkers form a unique macroscopic analogue of the pilot-wave theory imagined by \citet{deBroglie1987} as an interpretation of quantum mechanics. In a series of configurations recently summarised by \citet{Bush2015}, the behaviour of walkers is strongly reminiscent of that of quantum particles. Quantisation of bound states is observed in a central force field \citep{Perrard2014, Perrard2014c, Labousse2014, Labousse2016, Tambasco2016}, in a rotating frame \citep{Fort2010, Eddi2012, Harris2014, Oza2014, Oza2014b}, or with two-body interactions \citep{Protiere2006, Protiere2008, Borghesi2014, Filoux2015, Oza2017, Durey2017}. Walkers interact with submerged boundaries through non-specular reflection \citep{Pucci2016} and tunnelling \citep{Eddi2009, Carmigniani2014, Nachbin2017}. More remarkably, when they are confined in a cavity, they experience chaotic trajectories for which the corresponding statistics is close to the quantum statistics predicted by the Schr\"{o}dinger equation \citep{Harris2013, Gilet2014, Gilet2016}. They also seem to be diffracted when they pass one-by-one above a submerged slit \citep{Couder2006, Couder2012b, Dubertrand2016}, although it is not yet clear that the underlying statistics is really similar to that of a diffracted quantum particle \citep{Andersen2015}. Walkers therefore represent a unique opportunity to probe the boundary between the macroscopic world and the quantum realm. They allow to identify which specific behaviours of quantum particles could be explained or not through the deterministic coupling of a wave and a particle. 

An appropriate model of the damped Faraday waves emitted by the bouncing droplet is necessary to rationalise the dynamics of walkers. The first model has been proposed by \citet{Benjamin1954} for an inviscid fluid. Since there is no dissipation, the instability threshold is then identically zero. The Navier-Stokes equations simplify into a Matthieu equation for the interface elevation $\zeta(x,y,t)$. In the limit of infinite depth, it reads
\begin{equation}
\frac{d^2 \zeta}{dt^2} + \left[ g k + \frac{\sigma}{\rho} k^3 - \Gamma g k \cos (\Omega t) \right] \zeta = 0 \, ,
\end{equation}
where $k$ is the wavenumber. There is a subharmonic resonance for the Faraday wavenumber $k_F = 2\pi/\lambda_F$ such that:
\begin{equation}
\frac{\Omega^2}{4} = g k_F + \frac{\sigma}{\rho} k_F^3 \, .
\end{equation} 
In order to account for a finite viscosity $\nu$, a phenomenological linear damping term is added to this Matthieu equation (e.g. \cite{Kumar1994, Eddi2011, Molacek2013b, Blanchette2016}):
\begin{equation} \label{eq:DampedMatthieu}
\frac{d^2 \zeta}{dt^2} + 2 \alpha \nu k^2 \frac{d \zeta}{dt} + \left[ g k + \frac{\sigma}{\rho} k^3  - \Gamma g k \cos (\Omega t) \right] \zeta = 0 \, ,
\end{equation}
where $\alpha$ is a dimensionless factor. A slightly different model is derived by \citet{Milewski2015} from Navier-Stokes equations in the limit of small viscosity, without any additional phenomenological term  \citep{Durey2017}:
\begin{equation}  \label{eq:DampedMatthieuMilewski}
\frac{d^2 \zeta}{dt^2} + 4 \nu k^2 \frac{d \zeta}{dt} + \left[ g k + \frac{\sigma}{\rho} k^3 + 4 \nu^2 k^4 - \Gamma g k \cos (\Omega t) \right] \zeta = 0 \, .
\end{equation}
The wave field generated by one impact of the bouncing droplet at position ${\bf r}_n$ at time $t_{n}$ is then approximately \citep{Molacek2013b,Milewski2015}
\begin{equation} \label{eq:OldWaveField1impact}
\zeta_1({\bf r}-{\bf r}_n,t-t_{n}) = \frac{A \sin \Phi}{\sqrt{t-t_{n}}} \,J_0 \left( k_F |{\bf r} - {\bf r}_n| \right)\, \exp\left(- \frac{\Omega (t-t_{n})}{4\pi \Me}\right)\, \cos \left(\frac{\Omega t}{2}\right) \, ,
\end{equation}
where $\Me = \Omega\Gamma_F/[(8\pi \nu k_F^2)(\Gamma_F - \Gamma)]$ is defined as the memory parameter, $A$ is a constant amplitude that represents the strength of the impact, and $\Phi$ is linked (but not strictly equal) to the impact phase relative to the shaker. In the presence of obstacles, this equation \eqref{eq:OldWaveField1impact} can be replaced by a more elaborate Green's function that takes the wave reflection on boundaries into account \citep{Dubertrand2016}. It has also been suggested in \citep{Fort2010, Eddi2011} to add a spatial damping factor $e^{-|{\bf r} - {\bf r}_n| / l }$, where $l \sim 1.6 \lambda_F$ is a characteristic damping length. 

The wave field generated by a walker is assumed linear, so it can be computed by summing wave contributions from successive impacts:
\begin{equation} 
\zeta({\bf r},t) = \sum_n \zeta_1({\bf r}-{\bf r}_n,t-t_{n}) \, .
\end{equation}
A walker that does not experience any interaction with neither boundaries nor other walkers moves at constant horizontal speed and impacts the bath at regular times $t_{n} = t_i + 4 \pi n / \Omega$. This is referred as the ``stroboscopic approximation'' and it is implemented in all currently existing models of walkers, with the recent exception of \citet{Oza2017}. With the additional assumption that the horizontal translation of the walker during one rebound is much less than the Faraday wavelength $\lambda_F$, the sum can be replaced by a time integral, which yields an integro-differential equation for the walker trajectory \citep{Oza2013, Bush2014}. 

There are several issues to equation~\eqref{eq:OldWaveField1impact} and its variants. First, the elevation $\zeta$ of waves created by a single impact diverges in $t \rightarrow t_{n}$, which is unphysical. Second, the viscosity is taken into account through the damping factor $\alpha$ and the spatial damping length $l$, both of which are empirical corrections assumed to be valid only when viscous damping is weak. However, for the typical experimental conditions described above, the damping time $1/(4 \nu k_F^2) \sim 0.007$~s is actually smaller than the Faraday period $4\pi / \Omega = 0.025$~s. So viscous damping is definitely not weak, and one needs to check how much error is made by using these low-viscosity empirical approximations. Third, Faraday waves are assumed in phase with the forcing signal, and of constant amplitude $A$, without any proof nor experimental evidence. The dependence of $A$ to the phase shift between the impact and the forcing signal has never been evidenced experimentally. 

Without perturbations, a walker moves horizontally along a straight line, with a perfectly periodic bouncing motion. Therefore, these three limitations of equation~\eqref{eq:OldWaveField1impact} may induce discrepancies between models and experiments as soon as the walker's bouncing is perturbed, either by the presence of boundaries, other walkers, or waves that it left previously. Indeed, there is then no guarantee that the bouncing motion will remain periodic, and any perturbation of the impact phase would induce a different weight of each rebound to the wave field. This weight is also strongly dependent on temporal and spatial damping factors. The latter basically defines the distance below which a significant interaction between multiple walkers is expected. A wave field constructed with different weights for each impact will necessarily be qualitatively different, and so will the resulting walker trajectory be. The aforementioned perturbations by boundaries, other walkers and self-interferences, all correspond to configurations where quantum-like phenomena were observed. They seem to be key to the analogy between walkers and quantum particles.

% The limitations of equation~\eqref{eq:OldWaveField1impact} to accurately reproduce the perturbed wave field have motivated this work.
The main aim of this article is to obtain a quantitatively more accurate understanding of the surface waves emitted by walkers. The Faraday instability has been the subject of much theoretical work since \citet{Benjamin1954}. Most notably, \citet{Kumar1994} solved the full linearised Navier-Stokes equations (without any assumption on viscosity or depth) with a spectral method based on Floquet's theory. They showed, by comparing curves of marginal stability, that the damped Matthieu equation (\ref{eq:DampedMatthieu}) misses an important additional dissipation in the boundary layer at the liquid/air interface that only the full model can capture. However, because of the parametric forcing, the harmonics of the Floquet series are coupled to each other. Consequently, the dispersion relation can only be obtained by solving a discouraging eigenvalue problem for a matrix of infinite dimension. 

The truncation of the Floquet series of \citet{Kumar1994} to four harmonics ($\pm \Omega/2$ and $\pm 3\Omega/2$) has been investigated by \citet{Kumar1996}. With this truncated model, they could show that right above threshold ($\Gamma \gtrsim \Gamma_F$), Faraday waves are always subharmonic when both the depth $h$ and the Faraday wavelength $\lambda_F$ are much larger than the size of the boundary layer $\sqrt{2\nu/\Omega}$. This is the case for walkers since $\sqrt{2\nu/\Omega} \simeq 300 \mu$m $\ll h \simeq 6$~mm and $\lambda_F\simeq 4.7$~mm. Special wave patterns, including solitons, appear at the bicritical point where the boundary layer thickness is of the same order of magnitude as either the depth $h$ or the wavelength $\lambda_F$ \citep{Lioubashevski1996, Wagner2003}. An extreme truncation to only the subharmonic response $\pm \Omega/2$ has been performed by \citet{Muller1997}. They could then derive an analytical expression for the Faraday threshold $\Gamma_F$ in the limit of low viscosity (i.e. $\nu k^2 / \Omega \ll 1$) and large depth. 

In this paper, we follow the approach of \citet{Muller1997} to describe the Faraday waves emitted by walkers. In section \ref{sec:theory}, we first derive an analytical model of slightly-damped Faraday waves, while keeping control on the approximations that are made (e.g. the approximation of low viscosity). We then calculate the wave-field generated by a series of impacts. In section \ref{sec:Experiments}, we report the results of experiments on Faraday waves excited by an impacting bead. We show that our model goes beyond the identified limitations of equation (\ref{eq:OldWaveField1impact}) currently used to describe walkers. In section \ref{sec:Discussion}, we finally discuss the implication of this wave model for the development of the analogy between walkers and quantum particles.

\section{Surface waves theory for walkers}

\label{sec:theory}

\subsection{The surface wave equation}

\label{sec:sw}

\begin{figure}
\begin{psfrags}
\psfrag{zz}[r][r]{$\vec{e}_z$}\psfrag{y}[c][c]{$\vec{e}_y$}\psfrag{xx}[l][l]{$\vec{e}_x$}
\psfrag{z}[l][l]{$\zeta(\mathbf{r},t)$}\psfrag{g}[l][r]{\hspace{-0.5cm}$g(t)=g(1+\Gamma \cos\Omega t)$}
\psfrag{inf}[l][l]{$\infty$}\psfrag{rsnu}[l][l]{Liquid properties: $\rho$, $\sigma$, $\nu$}
\psfrag{press}[l][l]{Pressure field: $P(\vec{r},t)$}\psfrag{vit}[l][l]{Velocity field: $\vec{v}(\vec{r},t)$}
\includegraphics[width=0.8\textwidth]{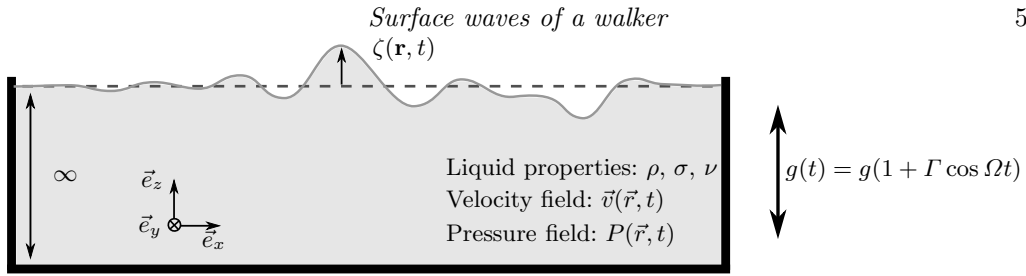}
\end{psfrags}
\caption{A liquid bath of infinite depth is vibrated vertically. The liquid experiences a time varying effective gravity $g(t)$. The elevation of the liquid surface is denoted $\zeta(x,y,t)= \zeta(\mathbf{r},t)$. The velocity and pressure fields inside the liquid bath are $\vec{v}(\vec{r},t)$ and $P(\vec{r},t)$, respectively.}
\label{fig:Schematics}
\end{figure}

We consider an infinitely extended and infinitely deep newtonian liquid with density $\rho$, surface tension $\sigma$, 
and viscosity $\nu$ (figure~\ref{fig:Schematics}). %For the case of interest (i.e. for walkers experiment) the liquid is silicone oil with the density $\rho = 956\,\mathrm{kg}/\mathrm{m}^3$, the surface tension $\sigma = 20.6 \times 10^{-3}\,\mathrm{N}/\mathrm{m}$, and the viscosity $\nu = 20 \times 10^{-6}\,\mathrm{m}^2/\mathrm{s}$. 
This liquid is incompressible, i.e. $\vec{\nabla}\cdot\vec{v}(\vec{r},t) = 0$, where $\vec{v}$ is the velocity
field within the liquid. It is subjected to a periodic vertical vibration of frequency $\Omega$ and acceleration amplitude $\Gamma g$, where $g = 9.81\,\mathrm{m}/\mathrm{s}^2$ is the gravity acceleration. Within the comoving frame, the liquid is then effectively experiencing a time-periodic modulation of gravity according to
\begin{equation}
g(t) = g \left[ 1 + \Gamma \cos(\Omega t) \right] \, . \label{eq:gt}
\end{equation}

The spatiotemporal evolution of the liquid is governed by the Navier-Stokes equation,
\begin{equation}
\frac{\partial}{\partial t} \vec{v}(\vec{r},t) + \left(\vec{v}(\vec{r},t)\cdot\vec{\nabla}\right) \vec{v}(\vec{r},t) = - \frac{1}{\rho} \vec{\nabla} P(\vec{r},t) - g(t) \vec{e}_z + \nu \vec{\nabla}^2 \vec{v}(\vec{r},t)\, , \label{eq:NS}
\end{equation}
where $\vec{r} = [x, y, z]$ and $P(\vec{r},t)$ denotes the pressure within the liquid relative to the atmospheric pressure. Without loss of generality, we choose a coordinate system whose origin is located at the surface of the liquid at rest, with the $z$-axis oriented upwards (i.e. opposite to the direction of the gravitational force). Considering small deviations from the stationary equilibrium state $\vec{v}^{(\mathrm{eq})}(\vec{r},t) = 0$ and $P^{(\mathrm{eq})}(\vec{r},t) = - \rho g(t) z$ of the liquid, we may linearise the Navier-Stokes equation \eqref{eq:NS}: 
\begin{equation}
\frac{\partial}{\partial t} \vec{v}(\vec{r},t) - \nu \vec{\nabla}^2 \vec{v}(\vec{r},t) = - \frac{1}{\rho} \vec{\nabla} p(\vec{r},t) \, ,\label{eq:ns}
\end{equation}
where $p(\vec{r},t) = P(\vec{r},t) + \rho g(t) z$ denotes the pressure perturbation. Owing to the incompressibility of the liquid, $p$ satisfies the Laplace equation $\vec{\nabla}^2 p(\vec{r},t) = 0$.

The boundary conditions at the surface $z=0$ of the liquid are also obtained under the assumption that the liquid is only weakly perturbed: the wave amplitude is assumed much smaller than the characteristic wavelength. The absence of tangential shear stress at the surface can then be expressed as
\begin{equation}
  \frac{\partial v_z}{\partial u} (x,y,0,t) + \frac{\partial v_u}{\partial z} (x,y,0,t) = 0 \, ,
  \label{eq:vb}
\end{equation}
with $u=x$ or $y$, which together with the incompressibility condition yields the boundary condition
\begin{equation}
  \frac{\partial^2 v_z}{\partial z^2} (x,y,0,t) = \left(\frac{\partial^2}{\partial x^2} + \frac{\partial^2}{\partial y^2} \right) v_z(x,y,0,t)
  \label{eq:vzb}
\end{equation}
for the vertical component of the liquid velocity. The pressure, on the other hand, satisfies the boundary condition
\begin{equation}
  P(x,y,0,t) = 2 \nu \rho \frac{\partial v_z}{\partial z} (x,y,0,t) - \sigma \left(\frac{\partial^2}{\partial x^2} + \frac{\partial^2}{\partial y^2} \right) \zeta(x,y,t) + P^{\rm ext}(x,y,t) \, , \label{eq:Pb}
\end{equation}
at the surface of the liquid. Here $\zeta(x,y,t)$ denotes the surface wave profile, i.e. the local elevation of the surface with respect to the equilibrium level $z=0$. It satisfies the kinematic condition
\begin{equation}
  \frac{\partial}{\partial t} \zeta(x,y,t) = v_z(x,y,0,t)\, . 
\end{equation}
The pressure $P^{\rm ext}(x,y,t)$ represents the spatiotemporal pressure profile that is induced from above the liquid, e.g., by the impact of a droplet on the surface \citep{Milewski2015}. This gives rise to the boundary condition
\begin{eqnarray}
  p(x,y,0,t) & = & 2 \nu \rho \frac{\partial v_z}{\partial z} (x,y,0,t) - \sigma \left(\frac{\partial^2}{\partial x^2} + \frac{\partial^2}{\partial y^2} \right) \zeta(x,y,t) + \rho g(t) \zeta(x,y,t) \nonumber \\
  && + P^{\rm ext}(x,y,t) \label{eq:pb}
\end{eqnarray}
for the local pressure perturbation.

Since the lateral extension of the bath is here supposed infinite, the above equations can be simplified by means of a Fourier ansatz defined through
\begin{equation}
X(x,y,z,t) = \frac{1}{2\pi} \int dk_x dk_y \, X_{\mathbf{k}}(z,t) \exp\left[i (k_x x + k_y y)\right] \label{eq:FTz}
\end{equation}
where $X$ is any of the variables $\zeta$, $p$, $v_z$, or $P^{\rm ext}$, and where the horizontal wave vector $\mathbf{k} = (k_x,k_y)$. Defining $k = |\mathbf{k}|$, the Laplace equation for the pressure perturbation is then expressed as
\begin{equation}
  \left( \frac{\partial^2}{\partial z^2} - k^2 \right) p_{\mathbf{k}}(z,t) = 0 \label{eq:pk} \, .
\end{equation}
This equation is straightforwardly solved as 
$p_{\mathbf{k}}(z,t) = p_{\mathbf{k}}(0,t) e^{kz}$ using the boundary conditions
\begin{equation}
  p_{\mathbf{k}}(0,t) = 2 \nu \rho \frac{\partial v_{z,\mathbf{k}}}{\partial z}(0,t) + \left[ \rho g(t) + \sigma k^2 \right] \zeta_{\mathbf{k}}(t) + P_{\mathbf{k}}^{\rm ext}(t)\quad\mathrm{and}\quad p_{\mathbf{k}}(-\infty ,t) =0\, .
\end{equation}
The Fourier transform of the linearised Navier-Stokes equation \eqref{eq:ns} then yields
\begin{eqnarray}
  \left( \frac{\partial}{\partial t} - \nu \frac{\partial^2}{\partial z^2}
  + \nu k^2 \right) v_{z,\mathbf{k}}(z,t) & = &
  - \left\{ 2 \nu \frac{\partial v_{z,\mathbf{k}}}{\partial z}(0,t)
  + \left[ g(t) + \frac{\sigma}{\rho} k^2 \right] \zeta_{\mathbf{k}}(t) \right\}
  k e^{kz} \nonumber \\ &&
  - \frac{k}{\rho} P_{\mathbf{k}}^{\rm ext}(t) e^{kz}\, ,  \label{eq:nsk}
\end{eqnarray}
with boundary conditions
\begin{equation}
\frac{\partial}{\partial t} \zeta_{\mathbf{k}}(t) = v_{z,\mathbf{k}}(0,t)
\label{eq:wxi}
\end{equation}
and 
\begin{equation}
  \frac{\partial^2 v_{z,\mathbf{k}}}{\partial z^2}(0,t) = - k^2 v_{z,\mathbf{k}}(0,t)\,.
  \label{eq:wbk}
\end{equation}

We now normalise the main variables with the characteristic timescale $2/\Omega$, and we define:
\begin{equation}
\tau = \frac{\Omega}{2} t, \quad w_{\mathbf{k}} =  \frac{2}{\Omega}v_{z,\mathbf{k}}, \quad \Pi_{\mathbf{k}} = \frac{4 k}{\Omega^2 \rho} P_{\mathbf{k}}^{\rm ext}, \quad \gamma_k = \frac{4\nu k^2}{\Omega}\, \quad \mathrm{and}\quad\omega_k^2 =\frac{4(gk+(\sigma/\rho)k^3)}{\Omega^2} \, .
\end{equation}
With this set of notations, equations~\eqref{eq:nsk} to \eqref{eq:wbk} become,  
%\begin{eqnarray}
%  \left( \frac{\partial}{\partial \tau} - \frac{2\nu}{\Omega} \frac{\partial^2}{\partial z^2}
%  + \frac{2\nu k^2}{\Omega} \right) w_{\mathbf{k}}(z,\tau) & = &
%  - \left\{ \frac{4 \nu}{\Omega} \frac{\partial  w_{\mathbf{k}}}{\partial z}(0,\tau)
%  + \frac{4}{\Omega^2}\left[ g(\tau) + \frac{\sigma}{\rho} k^2 \right] \zeta_{\mathbf{k}}(\tau) \right\}
%  k e^{kz} \nonumber \\ &&
%  - \Pi_k(\tau) e^{kz} \, \label{eq:nsk}
%\end{eqnarray}
\begin{align}
\nonumber\begin{split}
 \left( \frac{\partial}{\partial \tau} - \frac{\gamma_k}{2 k^2} \frac{\partial^2}{\partial z^2}
  + \frac{\gamma_k}{2} \right) w_{\mathbf{k}}(z,\tau)  = &
  - \left\{ \frac{\gamma_k}{k} \frac{\partial  w_{\mathbf{k}}}{\partial z}(0,\tau)
  + \left[\omega_k^2 +\frac{4\Gamma g k}{\Omega^2} \cos(2\tau) \right] \zeta_{\mathbf{k}}(\tau) \right\}
  e^{kz}  \\
  &  - \Pi_{\mathbf{k}}(\tau) e^{kz}\, ,
  \end{split}\\
\frac{\partial}{\partial \tau } \zeta_{\mathbf{k}}(\tau) = w_{\mathbf{k}}(0,\tau)&\quad\mathrm{and}\quad\frac{\partial^2 w_{\mathbf{k}}}{\partial z^2}(0,\tau) = - k^2 w_{\mathbf{k}}(0,\tau)\, .
\label{eq:nsk_new_notations}
\end{align}

We now define the Laplace transform on the dimensionless time $\tau$ with zero initial condition (bath surface at rest),
\begin{equation}
w_{\mathbf{k},s}(z) = \int_0^\infty w_{\mathbf{k}}(z,\tau) e^{-s\tau}\mathrm{d}\tau %\quad\mathrm{and}\quad w_k(z,\tau)=\frac{1}{2i\pi}\int_{-i\infty}^{i\infty} e^{s\tau} w_{k,s}(z) \mathrm{d}s\, ,
\end{equation}
and apply it to the three equations \eqref{eq:nsk_new_notations}: 
\begin{align}
\nonumber\begin{split}
 \left( s - \frac{\gamma_k}{2 k^2} \frac{\partial^2}{\partial z^2}
  + \frac{\gamma_k}{2} \right) w_{\mathbf{k},s}(z)  = &
  - \left\{ \frac{\gamma_k}{k} \frac{\partial  w_{\mathbf{k},s}}{\partial z}(0)
  + \omega_k^2 \zeta_{\mathbf{k},s}+\frac{2\Gamma g k}{\Omega^2}  \left(\zeta_{\mathbf{k},s+2i}+\zeta_{\mathbf{k},s-2i}\right)  \right\}
  e^{kz}  \\
  &  - \Pi_{\mathbf{k},s}\, e^{kz}\, ,
  \end{split}\\
s \zeta_{\mathbf{k},s} = w_{\mathbf{k},s}(0)&\quad\mathrm{and}\quad\frac{\partial^2 w_{\mathbf{k},s}}{\partial z^2}(0) = - k^2 w_{\mathbf{k},s}(0)\, .
\label{eq:nsk_new_notations_laplace}
\end{align}
The solution to the above differential equation can be expressed as
\begin{equation}
w_{\mathbf{k},s} = \left[-\gamma_k e^{\sqrt{1+2s/\gamma_k}\,kz}+(s+\gamma_k)e^{kz}\right]\zeta_{\mathbf{k},s}\, , 
\label{eq:vertical_velocity_solution}
\end{equation}
provided that $\zeta_{\mathbf{k},s}$ satisfies the relation %and leads to the equation of surface waves in Laplace domain,
\begin{equation}
f_k(s)\, \zeta_{\mathbf{k},s} + \frac{2\Gamma g k}{\Omega^2}\left(\zeta_{\mathbf{k},s+2i}+\zeta_{\mathbf{k},s-2i}\right) + \Pi_{\mathbf{k},s}=0\, ,
\label{eq:flf}
\end{equation}
where we define the function  
\begin{equation}
f_k(s) = (s+\gamma_k)^2 -\gamma_k^{3/2}\sqrt{\gamma_k+2s} + \omega_k^2 \,.
\label{eq:free_dispersion_relation}
\end{equation}
This function and all its derivatives are self-conjugated, i.e. $f_k(z^*) = (f_k(z))^*$. Similarly to \citet{Muller1997}, we finally obtain the equation for surface waves of a shaken bath in the time domain, by back Laplace-transforming the previous equation,
\begin{eqnarray}
\nonumber \left[\frac{\partial^2\cdot }{\partial \tau^2}+ 2\gamma_k \frac{\partial\cdot }{\partial \tau} + \gamma_k^2 + \omega^2_k\right]\zeta_{\mathbf{k}}(\tau) &-&\frac{\gamma_k^{3/2}}{\sqrt{2\pi}}\int_{-\infty}^\tau \frac{e^{-\gamma_k(\tau-u)/2}}{\sqrt{\tau-u}}\left(\gamma_k\zeta_{\mathbf{k}}(u)+2\frac{\partial \zeta_{\mathbf{k}}}{\partial u}(u)\right)\mathrm{d}u\\&+& \frac{4\Gamma g k}{\Omega^2}\cos({2 \tau})\zeta_{\mathbf{k}}(\tau) + \Pi_{\mathbf{k}}(\tau)=0\, ,
\label{eq:sws}
\end{eqnarray}
given that $\mathcal{L}^{-1}\left\lbrace 1/\sqrt{\gamma_k+2s}\right\rbrace = e^{-\gamma_k \tau /2 }/\sqrt{2\pi \tau}$. Equation~\eqref{eq:sws} can be integrated numerically to obtain the time dependency of each Fourier component $k$ of the surface wave elevation. Several terms of this equation are identical to terms of the equation \eqref{eq:DampedMatthieuMilewski} obtained by \citet{Milewski2015} in the limit of low viscosity. However, the convolution integral in equation~\eqref{eq:sws}, which is an additional signature of viscous shear, is missing in equation~\eqref{eq:DampedMatthieuMilewski}. 

\subsubsection*{Waves on an unshaken bath in the absence of external perturbation}

In the absence of any external perturbation of the liquid (i.e.\ for $\Gamma = 0$ and $\Pi_{\mathbf{k},s}= 0$ for all $\mathbf{k}$), the surface wave equation \eqref{eq:flf} becomes $f_k(s)\,\zeta_{\mathbf{k},s}=0$. The condition of existence of waves $\zeta_{\mathbf{k},s}\neq 0$ leads to the dispersion relation of free viscous gravity-capillary waves:
\begin{equation}
f_k(s) = (s+\gamma_k)^2 -\gamma_k^{3/2}\sqrt{\gamma_k+2s} + \omega_k^2 = 0\,.
\label{eq:free_dispersion_relation_null}
\end{equation}
Equation~\eqref{eq:free_dispersion_relation_null} can be solved for $s$ in the low-viscosity limit $\gamma_k\ll 1$:  
\begin{eqnarray}
 \nonumber \mathbb{R}(s) & = & -\gamma_k \left\{1 - \frac{1}{2} \left(\frac{\gamma_k}{\omega_k}\right)^{1/2} -\frac{1}{8} \left(\frac{\gamma_k}{\omega_k}\right)^{3/2}-\frac{1}{64} \left(\frac{\gamma_k}{\omega_k}\right)^{5/2}\right. \\
  &&\left.+  \mathcal{O}\left(\left(\frac{\gamma_k}{\omega_k}\right)^{9/2}\right) \right\}\,, \label{eq:gak} \\
 \nonumber \mathbb{I}(s) & = & \omega_k \left\{ 1 - \frac{1}{2} \left(\frac{\gamma_k}{\omega_k}\right)^{3/2} +\frac{1}{8} \left(\frac{\gamma_k}{\omega_k}\right)^{5/2}-\frac{1}{64} \left(\frac{\gamma_k}{\omega_k}\right)^{7/2}
  +\frac{1}{8} \left(\frac{\gamma_k}{\omega_k}\right)^{4} \right. \\
  &&\left. +  \mathcal{O}\left(\left(\frac{\gamma_k}{\omega_k}\right)^{9/2}\right) \right\}\,, \label{eq:omk}
\end{eqnarray}
%\begin{eqnarray}
% \mathcal{R}_e(s) & = & -\gamma_0 \left(1 - \frac{1}{2} \gamma_0^{1/2} -\frac{1}{8} \gamma_0^{3/2}-\frac{1}{64} \gamma_0^{5/2}
%  +  o\left(\gamma_0^{4}\right) \right)\,, \label{eq:gak} \\
%  \mathcal{I}_m(s) & = &  1 - \frac{1}{2} \gamma_0^{3/2} +\frac{1}{8} \gamma_0^{5/2}-\frac{1}{64} \gamma_0^{7/2}
%  +\frac{1}{8} \gamma_0^{4}  +  o\left(\gamma_0^{4}\right) \,, \label{eq:omk}
%\end{eqnarray}
where $\mathbb{R} (s)$ and $\mathbb{I}(s)$ represent respectively the damping and the frequency of the waves. The coefficients $\gamma_k$ and $\omega_k$ therefore correspond to the first order damping and frequency of gravity-capillary waves in the limit of low viscosity.

\subsection{Faraday waves, just below the Faraday threshold}

We now consider Faraday waves on a bath that is shaken just below the Faraday threshold $\Gamma_F$, i.e. in the conditions experienced by walkers. When these waves are subharmonic (as observed experimentally), they can be obtained by solving equation \eqref{eq:sws} with the Floquet ansatz 
\begin{equation}
\zeta_{\mathbf{k}}(t) = \sum_{l=-\infty}^\infty \zeta_{\mathbf{k}}^{(l)} e^{(2l - 1) i \tau} e^{\delta_k \tau} \, .
\label{eq:Ansatz_floquet}
\end{equation}
where $\delta_k$ is a small complex perturbation from the subharmonic resonance frequency that should vanish when the instability threshold is reached.  
%Since waves are damped below the instability threshold, their frequency $\omega$ should contain a finite (negative) imaginary part that corresponds to the decay rate of the wave.
Equivalently in the Laplace domain,
\begin{equation}
\zeta_{\mathbf{k},s} = \sum_{l=-\infty}^\infty \frac{\zeta_{\mathbf{k}}^{(l)}}{s + i (1 - 2 l) - \delta_k} \, .
\label{eq:Ansatz_floquetLaplace}
\end{equation}
Obtaining the exact solution to equation~\eqref{eq:flf} from this ansatz requires solving an infinite system of algebraic equations where all the Floquet components $\zeta_{\mathbf{k}}^{(l)}$ are coupled to each other \citep{Kumar1994}. As already suggested by \citet{Muller1997}, we can simplify this ansatz by restricting it to the two most relevant Floquet components $\zeta_{\mathbf{k}}^{(0)}$ and $\zeta_{\mathbf{k}}^{(1)}$ of the wave, namely those that are associated with angular frequencies in the vicinity of $\pm \Omega/2$. This truncated Floquet ansatz can be written as
\begin{eqnarray}
  \zeta_{\mathbf{k}}(\tau) &\simeq & \left(\zeta_{\mathbf{k}}^{(1)} e^{i \tau} + \zeta_{\mathbf{k}}^{(0)} e^{-i \tau}\right) e^{\delta_k \tau}\, , \nonumber \\ 
  \zeta_{\mathbf{k},s} &\simeq & \frac{\zeta_{\mathbf{k}}^{(1)}}{s - i - \delta_k} + \frac{\zeta_{\mathbf{k}}^{(0)}}{s + i - \delta_k}\, .
\label{eq:fq}
\end{eqnarray}
It is shown in Appendix~\ref{app:Floquet} that this truncation to the two first terms is valid as long as $|f_k(i)|^2/|f_k(3i)|^2 \ll 1$, which is true for wavenumbers $k \sim k_F$ in the low-viscosity limit. 

In the absence of external pressure applied on the bath, $\Pi_{\mathbf{k}}(\tau) = 0$, substitution of this ansatz into equation \eqref{eq:flf} and evaluation in $s=-i+\delta_k$ and $s=i+\delta_k$ yields
\begin{eqnarray}
f_k(i+\delta_k)\zeta_{\mathbf{k}}^{(1)}+\frac{2\Gamma gk}{\Omega^2}\zeta_{\mathbf{k}}^{(0)}&=&0 \label{eq:floquet1}\\
f_k(-i+\delta_k)\zeta_{\mathbf{k}}^{(0)}+\frac{2\Gamma gk}{\Omega^2}\zeta_{\mathbf{k}}^{(1)}&=&0 \label{eq:floquet2}\, .
\end{eqnarray}
These two equations admit nontrivial solutions if and only if
\begin{equation}
f_k(i+\delta_k)f_k(-i+\delta_k)-\left(\frac{2\Gamma g k}{\Omega^2}\right)^2  = 0 \,. \label{eq:cond}
\end{equation}

\subsubsection{Faraday instability}

The special choice $\delta_k = 0$ in equation~\eqref{eq:cond} yields the condition for the emergence of self-sustained Faraday waves at the subharmonic resonance. The dimensionless driving acceleration $\Gamma_k$ necessary to make the wave number $k$ unstable is then given by 
\begin{equation}
  \Gamma_k =\frac{\Omega^2|f_k(i)|}{2gk}\,  \label{eq:th}
\end{equation}
since $f_k(s)$ is self-conjugated. This function $\Gamma_k$ is represented in figure~\ref{fig:GammaDeltaTheta}a. There is always a minimum of $\Gamma_k$ that corresponds to the most unstable wave-number, $k_F$:
\begin{equation}
\left.\frac{\partial \Gamma_k}{\partial k}\right|_{k_F}=0 \,.
\label{eq:def_kf}
\end{equation}
The corresponding acceleration threshold is denoted $\Gamma_{F}=\Gamma_k (k_F)$. The effect of the truncation 
is seen in figure~\ref{fig:GammaDeltaTheta}a, where equation~\eqref{eq:th} is compared to the solution of the Floquet ansatz with a truncation to higher-order terms (Appendix A). Close to the Faraday threshold $k_F$, the difference between the two truncations becomes negligible.

\subsubsection{Waves just below the Faraday threshold}

When the driving acceleration $\Gamma$ is tuned slightly below the Faraday threshold $\Gamma_F$, some small complex detuning $\delta_k$ should arise for the wave numbers $k \sim k_F$, as prescribed by equation~\eqref{eq:cond}. The function $f_k(\pm i+\delta_k)$ can then be expanded to the second order in $\delta_k$:
\begin{equation} 
\nonumber f_k(\pm i+\delta_k)= f_k(\pm i)+ \delta_k\dot f_k(\pm i)  + \delta_k^2\ddot f_k(\pm i)/2\,,
\end{equation} 
where $\dot f_k = \partial f_k/\partial s$. Substitution in equation~\eqref{eq:cond} yields
\begin{eqnarray}
 \nonumber \frac{\delta_k^2}{2}\left(f_k(i)\ddot{f}_k(-i)+f_k(-i)\ddot{f}_k(i) + 2\dot{f}_k(i)\dot{f}_k(-i)\right) &+&\\ \delta_k\left(f_k(i)\dot{f}_k(-i)+f_k(-i)\dot{f}_k(i)\right) &+& \left(\frac{2 g k}{\Omega^2}\right)^2 (\Gamma_{k}^2 -\Gamma^2)=0 \, . \label{eq:d2}
\end{eqnarray}
%We recall that, 
% \begin{eqnarray}
%  f_k(s) &=& (s+\gamma_k)^2+\omega_k^2 - \gamma_k^{3/2}\sqrt{\gamma_k+2s}\,, \\ 
%  \dot f_k(s) &=& 2(s+\gamma_k)-\frac{\gamma_k^{3/2}}{\sqrt{\gamma_k+2s}}\,, \\
%  \ddot f_k(s) &=& 2 +\left(\frac{\gamma_k}{\gamma_k+2s}\right)^{3/2}\,.
% \end{eqnarray}
%These function have the properties of being self conjugated, $f_k(s^*) = f_k(s)^*$, $\dot f_k(s^*) = \dot f_k(s)^*$ and  $\ddot f_k(s^*) =\ddot f_k(s)^*$. Note also that if $(z_1, z_2)\in \mathbb{C}$, then $z_1\,z_2^* + z_1^*\,z_2 = 2\left(\mathcal{R}(z_1)\mathcal{R}(z_2)+\mathcal{I}(z_1)\mathcal{I}(z_2)\right)$.
The coefficients of this second order polynomial in $\delta_k$ are all real, because $f_k$ and its derivatives are self-conjugated. If the corresponding discriminant is positive, i.e. if 
\begin{equation}
\Gamma_{k}^2 -\Gamma^2<
\left(\frac{\Omega^2}{2 g k}\right)^2\frac{\left(f_k(i)\dot{f}_k(-i)+f_k(-i)\dot{f}_k(i)\right)^2}{2\left(f_k(i)\ddot{f}_k(-i)+f_k(-i)\ddot{f}_k(i) + 2\dot{f}_k(i)\dot{f}_k(-i)\right)}\, ,
\label{eq:FaradayWindow}
\end{equation}
then both solutions $\delta_k \in \mathbb{R}$ and the resulting waves (cf. equation~\ref{eq:fq}) are locked at half the driving frequency, i.e. at $\Omega/2$. For a given $\Gamma \lesssim \Gamma_F$, equation~\eqref{eq:FaradayWindow} is valid for a small range of $k$ around $k_F$ that we call the Faraday window. Outside this range, waves are travelling waves with group velocity different from 0, and their frequency is detuned from the forcing frequency since the corresponding $\delta_k$ is complex. The latter waves compound the capillary wave packet by opposition to standing Faraday waves.

%leading to the condition on $k$, at the smallest order in $\gamma_k$, see detailed expansions in Appendix,
%\begin{equation} 
% \Gamma_k^2 - \Gamma^2<\left(\frac{\Omega^2}{gk}\right)^2\left((\omega_k^2+1)\gamma_k^2 -\frac{\omega_k^2+3}{2}\gamma_k^{5/2}+\mathcal{O}(\gamma_k^3)\right)
%\end{equation}

\begin{figure}
\begin{psfrags}
\psfrag{a}[c][c]{a.}\psfrag{b}[c][c]{b.}\psfrag{c}[c][c]{c.}\psfrag{d}[c][c]{d.}
\psfrag{f}[c][c]{Faraday}\psfrag{w}[c][c]{window}
\psfrag{fw}[c][c]{}\psfrag{dk}[c][c]{$\mathbb{R}(\delta_k)$}
%\psfrag{dk1}[l][r]{$\delta_k^+$}\psfrag{dk2}[l][r]{$\delta_k^-$}
\psfrag{gk}[c][c]{$\Gamma_k$}\psfrag{k}[c][c]{$k$ (m$^{-1}$)}
\psfrag{kF}[c][c]{$k_F$}\psfrag{gF}[c][c]{$\Gamma_F$}
\psfrag{Im}[l][l]{$\mathbb{I}(i+\delta_k)$}
\psfrag{Im1}[l][l][0.8]{$\mathbb{I}(i+\delta_k^+)$}
\psfrag{Im2}[c][r][0.8]{$\mathbb{I}(i+\delta_k^-)$}
\psfrag{4}[cr][cr][0.8]{4}\psfrag{6}[cr][cr][0.8]{6}\psfrag{8}[cr][cr][0.8]{8}\psfrag{5}[cr][cr][0.8]{5}\psfrag{7}[cr][cr][0.8]{7}
\psfrag{1000}[ct][cb][0.8]{1000}\psfrag{1200}[cb][ct][0.8]{1200}\psfrag{1400}[c][c][0.8]{1400}\psfrag{1600}[c][c][0.8]{1600}\psfrag{1800}[c][c][0.8]{1800}
\psfrag{-0.4}[cr][cr][0.8]{-0.4}\psfrag{-0.3}[cr][cr][0.8]{-0.3}\psfrag{-0.2}[cr][cr][0.8]{-0.2}\psfrag{-0.1}[cr][cr][0.8]{-0.1}\psfrag{0}[cr][cr][0.8]{0}\psfrag{0.6}[rb][cr][0.8]{0.6}\psfrag{0.8}[cr][cr][0.8]{0.8}\psfrag{1}[cr][cr][0.8]{1}\psfrag{1.2}[cr][cr][0.8]{1.2}\psfrag{1.4}[cr][cr][0.8]{1.4}
\psfrag{f}[c][c]{Faraday}\psfrag{w}[c][c]{window}
\psfrag{fw}[c][c]{}\psfrag{dk}[l][l]{$\mathbb{R}(\delta_k)$}
\psfrag{dk1}[l][r][0.8]{$\delta_k^+$}\psfrag{dk2}[l][r][0.8]{$\delta_k^-$}
\psfrag{dF}[c][c]{$\delta_F$}
\psfrag{gk}[c][c]{$\Gamma_k$}\psfrag{k}[c][c]{$k$ (m$^{-1}$)}
\psfrag{kF}[c][c]{$k_F$}\psfrag{tF}[r][l]{$\theta_F$}
\psfrag{tk}[c][c]{$\theta_k$}
\psfrag{t1}[c][c][0.8]{$\theta_k^+$}\psfrag{t2}[c][c][0.8]{$\theta_k^-$}
\psfrag{p2}[r][r][0.8]{${\pi/2}$}\psfrag{p4}[r][r][0.8]{${=\pi/4}$}\psfrag{-p2}[rb][r][0.8]{$-{\pi/2}$}\psfrag{-p4}[r][r][0.8]{-${\pi/4}$}
\psfrag{1000}[ct][cb][0.8]{1000}\psfrag{1200}[ct][cb][0.8]{1200}\psfrag{1400}[ct][cb][0.8]{1400}\psfrag{1600}[ct][cb][0.8]{1600}\psfrag{1800}[ct][cb][0.8]{1800}
\psfrag{-0.4}[cr][cr][0.8]{-0.4}\psfrag{-0.3}[cr][cr][0.8]{-0.3}\psfrag{-0.2}[cr][cr][0.8]{-0.2}\psfrag{-0.1}[cr][cr][0.8]{-0.1}\psfrag{0}[cr][cr][0.8]{0}
 \includegraphics[width=\textwidth]{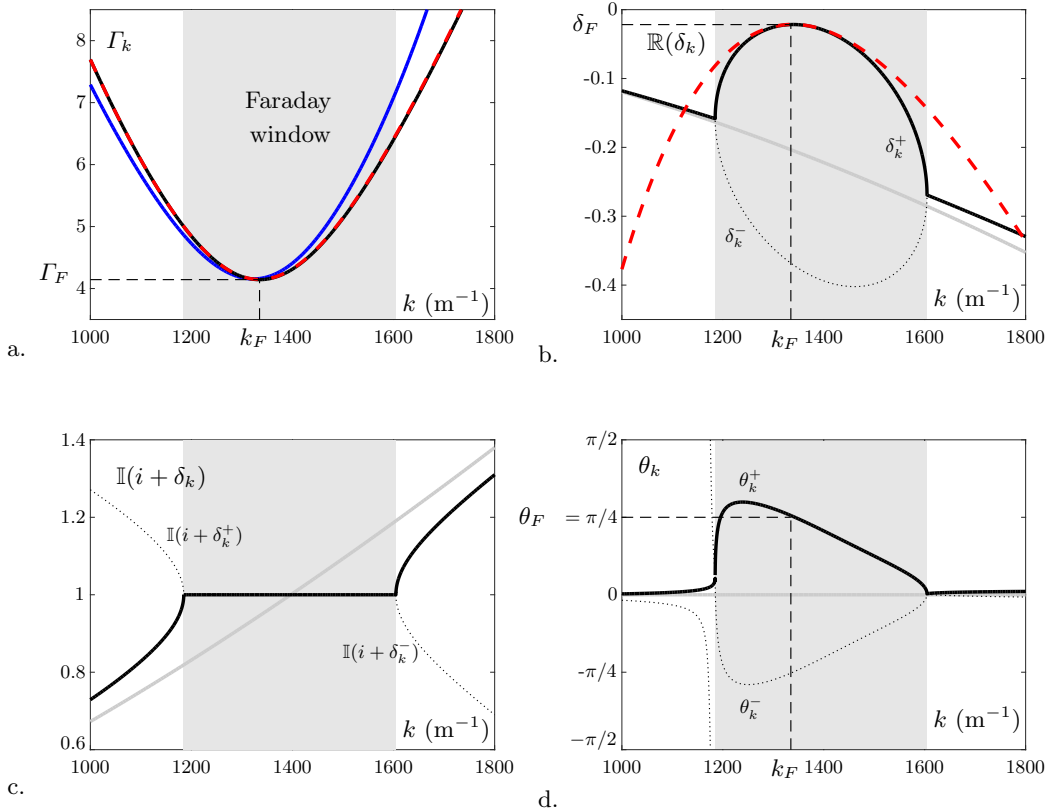}
\end{psfrags}
\caption{Color online.
\textit{a.} Acceleration threshold $\Gamma_k$ required to sustain waves of wavenumber $k$. The minimum of $\Gamma_k$ gives the most unstable wave number, $k_F$. \textit{b.} Damping of the Faraday waves $\mathbb{R}(\delta_k)$. \textit{c.} Frequency of the Faraday wave $\mathbb{I}(i+\delta_k)$. \textit{d.} Phase shift $\theta_k$ of the Faraday waves with respect to the forcing signal. In \textit{a}, the blue curve represents the solution of the Floquet's ansatz with truncation to 4 harmonics (equation~\eqref{eq:exact_sol_floquet2}), while the black curve corresponds to the truncation to 2 harmonics (equation~\eqref{eq:th}). In \textit{b, c}, the dotted line and the thick solid line represent equations \eqref{eq:largest_damping_k} and \eqref{eq:smallest_damping_k}, respectively. In \textit{d}, they represent equation~\eqref{eq:thetas_def}. In \textit{a, b}, the red dashed curve is the low-viscosity approximation of the truncation to two harmonics (equations~\eqref{eq:estimatefi2} and \eqref{eq:damping_k_approx}).
In \textit{b-c}, the damping and frequency of unforced waves (equations~\eqref{eq:gak} and \eqref{eq:omk}, respectively) are represented in light grey. In \textit{a-d}, all equations have been evaluated with numerical values of table \ref{tab:notations}. In particular, the distance to threshold $\mathcal{M} = (\Gamma_F - \Gamma)/\Gamma_F = 10$. The grey shade indicates the Faraday window, i.e. the range of wavenumbers $k$ that produce waves locked to half the driving frequency.}
\label{fig:GammaDeltaTheta}
\end{figure}

%Taylor series of the coefficients of equation~\eqref{eq:d2} in the limit of low-viscosity (in Appendix~\ref{app:low_viscosity}) indicate that all these coefficients are are positive indicating two real negative roots. 
%Around $k_F$, all wave numbers are locked in but are damped with a damping factor that depends on $k$.

Equation~\eqref{eq:d2} admits two solutions
\begin{equation}
\delta^{\pm}_k = -\frac{b}{2a}\left(1\mp\sqrt{1- \frac{4ac}{b^2}(\Gamma_k^2-\Gamma^2)}\right)
\end{equation}
where coefficients $a$, $b$ and $c$ are obtained by identification of equation~\eqref{eq:d2} with $a\delta_k^2 + b\delta_k+c(\Gamma_k^2-\Gamma^2) =0$. Since $\Gamma_k \simeq \Gamma$ inside the Faraday window, these solutions can be expanded in Taylor series for $| 4 a c (\Gamma_k^2-\Gamma^2) / b^2 | \ll 1$. To the first order, 
\begin{equation}
\delta_k^{-} = -\frac{b}{a} \left[ 1 + \mathcal{O} \left(\frac{a c (\Gamma_k^2-\Gamma^2)}{b^2} \right) \right]
\simeq -\frac{2\left(f_k(i)\dot{f}_k(-i)+f_k(-i)\dot{f}_k(i)\right)}{f_k(i)\ddot{f}_k(-i)+f_k(-i)\ddot{f}_k(i) + 2\dot{f}_k(i)\dot{f}_k(-i)}\, ,
\label{eq:largest_damping_k}
\end{equation}
and
\begin{equation}
\delta_k^{+} = -\frac{c  (\Gamma_k^2-\Gamma^2)}{b} \left[ 1 + \mathcal{O} \left(\frac{a c (\Gamma_k^2-\Gamma^2)}{b^2} \right) \right] \simeq -\frac{(2 g k/\Omega^2)^2 (\Gamma_{k}^2 -\Gamma^2)}{f_k(i)\dot{f}_k(-i)+f_k(-i)\dot{f}_k(i)}\, .
\label{eq:smallest_damping_k}
\end{equation}
The real and imaginary parts of these solutions $\delta_k^{\pm}$ are represented in figure~\ref{fig:GammaDeltaTheta}b and \ref{fig:GammaDeltaTheta}c, respectively. The imaginary part $\mathbb{I}(\delta_k^{\pm})$ corresponds to some detuning from the Faraday frequency $\Omega/2$, and it is identically 0 in the Faraday window around $k_F$. The corresponding waves are standing. Outside from this Faraday window, $\mathbb{I}(\delta_k^{\pm}+i)$ increases with $k$, so the group velocity is positive and the solution corresponds to travelling waves. Inside the Faraday window, the real part of $\delta_k^{-}$ remains largely negative so it describes Faraday waves that are strongly damped and that vanish quickly. By contrast, when $\Gamma$ approaches $\Gamma_F$, the real part of $\delta_k^{+}$ can be arbitrarily close to 0 for $k \simeq k_F$, in the Faraday window. Consequently, $\delta_k^{+}$ describes some long-lived standing Faraday waves. These latter waves are those observed to travel with the walker, and they are crucial to the description of its dynamics since their amplitude remains significant during many rebounds. 

The memory $\Me$ of the waves, initially described by \citet{Eddi2011}, is defined as the damping time of these long-lived Faraday waves, expressed in number of rebounds of the droplet. Since the bouncing period corresponds to the Faraday period $T_F = 4 \pi / \Omega$, the memory is given by
\begin{equation}
\Me =  \frac{-1}{2\pi \delta_F} \simeq \frac{1}{4\pi} \frac{f_F(i) \dot{f}_F(-i) + f_F(-i) \dot{f}_F(i)}{|f_F(i)|^2} \mathcal{M} \, .
\label{eq:definition_memory}
\end{equation}
where $\delta_F = \lim_{k \rightarrow k_F} \mathbb{R} (\delta_k^+)$ and where
\begin{equation}
\mathcal{M} = \frac{\Gamma_F}{\Gamma_F - \Gamma}
\end{equation}
is called the "distance to the Faraday threshold". This latter parameter has often been used as a proxy for the memory in experimental studies of walkers \citep{Eddi2011}. 

For $\Gamma \lesssim \Gamma_F$ (i.e. for $\mathcal{M} \gg 1$), we can expand $\delta_k^{+}$ in a Taylor series around $k_F$. Recalling that $\partial \Gamma_k/\partial k = 0$ in $k=k_F$, we obtain 
\begin{eqnarray}
\nonumber \delta_k^{+} &=& -\left[\frac{\Gamma_F(2 g k_F/\Omega^2)^2}{\left(f_F(i)\dot{f}_F(-i)+f_F(-i)\dot{f}_F(i)\right)}\left(\frac{\partial^2\Gamma_{k}}{\partial k^2}\right)_F(k-k_F)^2+\mathcal{O}\left((k-k_F)^3\right)\right]\\&-& \frac{2}{\mathcal{M}}\left[\frac{\Gamma_F^2(2 g k_F/\Omega^2)^2}{f_F(i)\dot{f}_F(-i)+f_F(-i)\dot{f}_F(i)}+\mathcal{O}(k-k_F)\right]+\mathcal{O}\left(\mathcal{M}^{-2}\right).
\label{eq:delta+dev}
\end{eqnarray}
where $f_F= f_{k_F}$ and $\dot{f}_F= \dot{f}_{k_F}$. We identify this expression with $\delta_k^+ = \delta_F - D ( k - k_F )^2 + \mathcal{O}\left((k-k_F)^3\right) + \mathcal{O}(\mathcal{M}^{-1})$ and define the diffusion coefficient $D$ as
\begin{equation}
  D = -\left.\frac{1}{2} \frac{\partial^2\delta_k^+}{\partial k^2} \right|_{k=k_F} = \frac{\Gamma_F(2 g k_F/\Omega^2)^2}{\left(f_F(i)\dot{f}_F(-i)+f_F(-i)\dot{f}_F(i)\right)}\left(\frac{\partial^2\Gamma_{k}}{\partial k^2}\right)_F\,
\label{eq:D}
\end{equation}
Since
\begin{equation}
\left(\frac{\partial^2\Gamma_{k}}{\partial k^2}\right)_F = \frac{1}{2k_F^2\Gamma_F}\left.\frac{\partial^2\left(k\Gamma_k\right)^2}{\partial k^2}\right|_{F} - \frac{\Gamma_F}{k_F^2}
\end{equation} 
and $k^2\Gamma_k^2 = \Omega^4|f_k(i)|^2/(2g)^2$, $D$ can be expressed in terms of $f_k$ as
\begin{equation}
D=\frac{(\partial^2|f_k(i)|^2/\partial k^2)_F-2|f_F(i)|^2/k_F^2}{2\left(f_F(i)\dot{f}_F(-i)+f_F(-i)\dot{f}_F(i)\right)}\, .
\label{eq:D_exact}
\end{equation} 

The amplitude $\zeta_k^+$ of the long-lived Faraday waves (respectively $\zeta_k^-$ for short-lived waves) is given from \eqref{eq:floquet1}:
%\begin{equation}
%\zeta_k^+(\tau) \propto e^{\delta_k^+ \tau}\left(e^{i\tau}-\frac{f_k(i+\delta_k^+)}{2\Gamma g k/\Omega^2} e^{-i\tau}\right),\,\, \mathrm{and}\,\,\zeta_k^-(\tau) \propto e^{\delta_k^- \tau}\left(e^{i\tau}-\frac{f_k(i+\delta_k^-)}{2\Gamma g k/\Omega^2} e^{-i\tau}\right)\, .
%\end{equation}
\begin{equation}
\zeta_k^\pm(\tau) \propto e^{\delta_k^\pm \tau}\left(e^{i\tau}-\frac{f_k(i+\delta_k^\pm)}{2\Gamma g k/\Omega^2} e^{-i\tau}\right)\, .
\end{equation}
We introduce
\begin{equation}
\theta_k^\pm=\mathrm{arctan}\left(\frac{\mathbb{I} \left\lbrace f_k(i+\delta_k^\pm)\right\rbrace}{2\Gamma gk /\Omega^2-\mathbb{R} \left\lbrace f_k(i+\delta_k^\pm)\right\rbrace}\right)\,,
\label{eq:thetas_def}
\end{equation}
in order to express the total wave field $\zeta_k(\tau)$ as
\begin{equation}
\zeta_k(\tau) = A_k^+ e^{\delta_k^+ \tau}\cos\left(\tau+ \theta_k^+\right)  +A_k^{-}e^{\delta_k^- \tau}\cos\left(\tau+ \theta_k^-\right)
\label{eq:allFaraday}
\end{equation} 
where $A_k^+$ and $A_k^-$ stand for the relative amplitude of long-lived and short-lived waves, respectively. The angles $\theta_k^{\pm}$ then correspond to the phase shift between these waves and the driving acceleration. They are represented in figure~\ref{fig:GammaDeltaTheta}d. Outside the Faraday window, they are both close to 0, while inside the Faraday window they strongly vary with $k$ in a non-trivial way. Remarkably around $k_F$, these phase shifts are $\theta_F^{\pm} \simeq \pm \pi/4$.

\subsection{Single impact of a droplet}
\label{sec:ImpulseResponse}

We now consider the impact of a droplet at the surface of the liquid at a time $\tau_i$. While sophisticated theoretical frameworks have been developed in order to accurately calculate the interaction of a droplet with the surface
of a liquid \citep{Molacek2012, Molacek2013a, Milewski2015}, we shall, for illustration purpose, employ a rather 
simplified description of the droplet impact. Namely, we model its temporal profile with a delta function. The corresponding dimensionless pressure exerted by the droplet on the liquid bath is then
%We furthermore assume that the droplet has perfect axisymetric shape and impinges upon the surface at $x=y=0$.
\begin{equation}
  \Pi_{\mathbf{k}}(\tau) = \delta(\tau-\tau_i) v_{\mathbf{k}}  \,. \label{eq:P}
\end{equation}

By integrating equation \eqref{eq:sws} across the delta kick, it is shown that $v_k$ corresponds to the negative change of velocity of $\zeta_{\mathbf{k}}$ following impact. If the surface is perfectly flat and at rest before the impact, the wave profile is axisymmetric, i.e. $\zeta_{\mathbf{k}}(\tau) = \zeta_{k}(\tau)$. The initial conditions in $\tau \rightarrow \tau_i^+$ are 
$\zeta_k = 0$ and $\partial \zeta_k /\partial \tau= - v_k$. They determine the coefficients $A_k^{\pm}$ of equation~\eqref{eq:allFaraday}:
%If $k$ lies within the Faraday window defined by the inequality 
%\eqref{eq:prox}, for instance, we obtain
%\begin{equation}
%  A^\pm = - \frac{ v_k^\circ\, e^{-\delta_k^\pm t_\circ}}{\cos(t_\circ+\theta^\pm)\left((\delta_k^\pm - \delta_k^\mp)+\tan(t_\circ+\theta_k^\mp)-\tan(t_\circ+\theta_k^\pm)\right)} \,.
%\end{equation}
%\begin{equation}
%  A^\pm = - \frac{2 \cos(t_\circ+\theta_k\mp) v_k^\circ\, e^{-\delta_k^\pm t_\circ}}{\left(\delta_k^\pm - \delta_k^\mp\right)\left(\cos(2t_\circ+\theta_k^\pm+\theta_k^{\mp})+\cos(\theta_k^\pm-\theta_k^\mp)\right)-2\sin(\theta_k^\pm-\theta_k^\mp)} \,.
%\end{equation}
\begin{equation}
A_k^\pm = - \frac{2 \cos(\tau_i+\theta_k^\mp) v_k\, e^{-\delta_k^\pm \tau_i}}{\left(\delta_k^\pm - \delta_k^\mp\right)\left(\cos(2\tau_i+\theta_k^\pm+\theta_k^{\mp})+\cos(\theta_k^\pm-\theta_k^\mp)\right)-2\sin(\theta_k^\pm-\theta_k^\mp)} \,.
\label{eq:ThAmplitude}
\end{equation}

Transforming back to the spatial domain according to equation~\eqref{eq:FTz} then yields the radially symmetric surface wave profile
\begin{equation}
\zeta_1(r,\tau) = \int_0^\infty \zeta_k(\tau) J_0(kr) k dk 
\label{eq:zeta}
\end{equation}
that is generated by a delta kick, where $J_0$ is the Bessel function of the first kind of order zero. 

At large times $\tau \gg \tau_i$ for $\Gamma \lesssim \Gamma_F$, the wave field is dominated by the contribution $\zeta_k^+$, so 
\begin{equation}
\zeta_1(r,\tau) = \int_0^\infty B_k^+(\tau_i)\, \cos (\tau + \theta_k^+) e^{\delta_k^+ (\tau - \tau_i)} J_0(kr) k dk \, ,
\label{eq:zeta1}
\end{equation}
where $B_k^+(\tau_i) = A_k^+ e^{\delta_k^+ \tau_i}$. With the expansion $\delta_k^+ = \delta_F - D ( k - k_F )^2 +\mathcal{O}\left( (k-k_F)^3\right)+ \mathcal{O}(\mathcal{M}^{-1})$, we can then approximate this integral thanks to the saddle-point method in the formal limit $\tau \to \infty$. Calculations detailed in appendix~\ref{app:hankel} yield:
\begin{equation}
\zeta_1(r,\tau) \simeq  B_F^+(\tau_i) k_F \sqrt{\frac{\pi}{D(\tau-\tau_i)}} \cos (\tau + \theta_F^+) 
 J_0(k_F r) \exp\left[ - \frac{\tau - \tau_i}{2 \pi \Me} - \frac{r^2}{4 D (\tau - \tau_i)} \right] \, ,
\label{eq:zeta2}
\end{equation}
which is at least valid when $k_F r \ll 1$ and $k_F r \gg 1$. Equation~\eqref{eq:zeta2} is the impulse response of the Faraday waves in the long-time limit. According to Duhamel's principle, the response to any sophisticated impact signal $\Pi_k(\tau)$ can be inferred by convoluting this impulse response with $\Pi_k(\tau)$. 

Equation~\eqref{eq:zeta2} may be compared to the main previous model of waves generated by a single impact of a droplet (equation~\ref{eq:OldWaveField1impact}), which can be rewritten as:
\begin{equation}
\zeta (r,t) = \frac{A(\tau_i)}{\sqrt{\tau-\tau_i}} \cos \left( \tau \right) J_0 \left( k_F r \right) \exp \left(- \frac{\tau - \tau_i}{2\pi \Me}\right) 
\label{eq:OldWaveField1impact2}
\end{equation}
The main additions of equation~\eqref{eq:zeta2} to equation~\eqref{eq:OldWaveField1impact2} are \textit{(i)} the phase shift $\theta_F^+$, \textit{(ii)} a diffusive spreading in the exponential decay (diffusion coefficient $D$), and \textit{(iii)} a well-defined amplitude $B_F^+$ that depends on the impact phase $\tau_i$ through equation~\eqref{eq:ThAmplitude}.

\subsection{Surface waves emitted by a walker}\label{sec:Static_bouncer}

The long-lived Faraday wave generated by a walker is obtained by superimposing impulse responses corresponding to each droplet impact at time $\tau_n$ and position $\mathbf{r}_n$. In the absence of other walkers, submerged boundaries or body forces, a walker bounces periodically every Faraday period and moves in a straight line at constant speed $V_w$. 
Consequently,
\begin{equation}
\tau_n = \tau_i + 2 \pi n, \quad n \in \mathbb{Z}, \mbox{ and } \mathbf{r}_n = -(\tau - \tau_n) \mathbf{v}
\end{equation}
where $v = 2 V_w / \Omega$ is the normalised walking speed. The origin $\mathbf{r} = 0$ is set where the walker is at time $\tau$. The Faraday wave field is then given by
\begin{equation}
  \zeta_w(\mathbf{r},\tau) = \sum_{n = -\infty}^{[(\tau - \tau_i/2\pi]} \zeta_1(|\mathbf{r}-\mathbf{r}_n|,\tau - 2\pi n)
  \label{eq:accFaraday}
\end{equation}
where $[(\tau - \tau_i)/2\pi]$ denotes the largest integer smaller than $(\tau - \tau_i)/2\pi$.

\begin{figure}
\begin{center}
\begin{psfrags}
\psfrag{0}[c][c]{0}\psfrag{1}[c][c]{1}\psfrag{20}[c][c]{20}\psfrag{40}[c][c]{40}\psfrag{60}[c][c]{60}
\psfrag{a0}[c][c][0.7]{0}\psfrag{5}[c][c][0.7]{5}\psfrag{15}[c][c][0.7]{15}\psfrag{10}[c][c][0.7]{10}
\psfrag{bouncer}[c][c]{bouncer}\psfrag{walker}[c][c]{walker}
\psfrag{25}[c][c][0.7]{25}\psfrag{50}[c][c][0.7]{50}\psfrag{75}[c][c][0.7]{75}
\psfrag{1000}[c][c]{1000}\psfrag{2000}[c][c]{2000}\psfrag{3000}[c][c]{3000}\psfrag{4000}[c][c]{4000}
\psfrag{e0}[r][r][0.95]{1}\psfrag{e-2}[r][r][0.95]{$10^{-2}$}\psfrag{e-4}[r][r][0.95]{$10^{-4}$}\psfrag{e-6}[r][r][0.95]{$10^{-6}$}
\psfrag{0.01}[c][c]{0.01}\psfrag{0.03}[c][c]{0.03}\psfrag{0.05}[c][c]{0.05}\psfrag{0.02}[c][c]{0.02}\psfrag{0.04}[c][c]{0.04}\psfrag{-0.05}[c][c]{-0.05}\psfrag{p}[c][c]{1 cm}
\psfrag{p}[c][c]{$\pi/2$}\psfrag{2p}[c][c]{$\pi$}\psfrag{3p}[c][c]{$3\pi/2$}\psfrag{4p}[c][c]{$2\pi$}
\psfrag{a}[ct][cb]{a}\psfrag{b}[ct][cb]{b}\psfrag{dl}[cb][ct]{Damping length, $l /\lambda_f$}
\psfrag{time}[c][c]{time (s)}\psfrag{distance}[ct][cb]{Distance from walker (mm)}\psfrag{Rwg}[cb][ct]{Radial wave gradient}\psfrag{Ndt}[ct][cb]{Normalized time $\tau/\pi$}\psfrag{Ndr}[c][c]{Normalized radius $r/\lambda_f$}
%\psfrag{v0}[l][l]{2 mm/s}\psfrag{v1}[l][l]{8.5 mm/s}\psfrag{v2}[c][c]{15 mm/s}
\psfrag{v0}[l][l][0.9]{0}\psfrag{v1}[l][l][0.9]{7.5}\psfrag{v2}[l][l][0.9]{15}\psfrag{v}[cb][rt][0.9]{$v$ (mm/s)}
\psfrag{Spatial}[c][c]{}\psfrag{ip}[ct][cb]{Impact phase $\phi$}\psfrag{NA}[cb][ct]{Normalized amplitude}\psfrag{Memory}[ct][cb]{Distance to threshold, $\mathcal{M}$}
\psfrag{Memory1}[c][c][0.7]{$\mathcal{M}$}\psfrag{dl1}[c][c][0.7]{$l /\lambda_f$}
\textit{a.}\hspace{2mm} 
\includegraphics[width=0.23\textwidth]{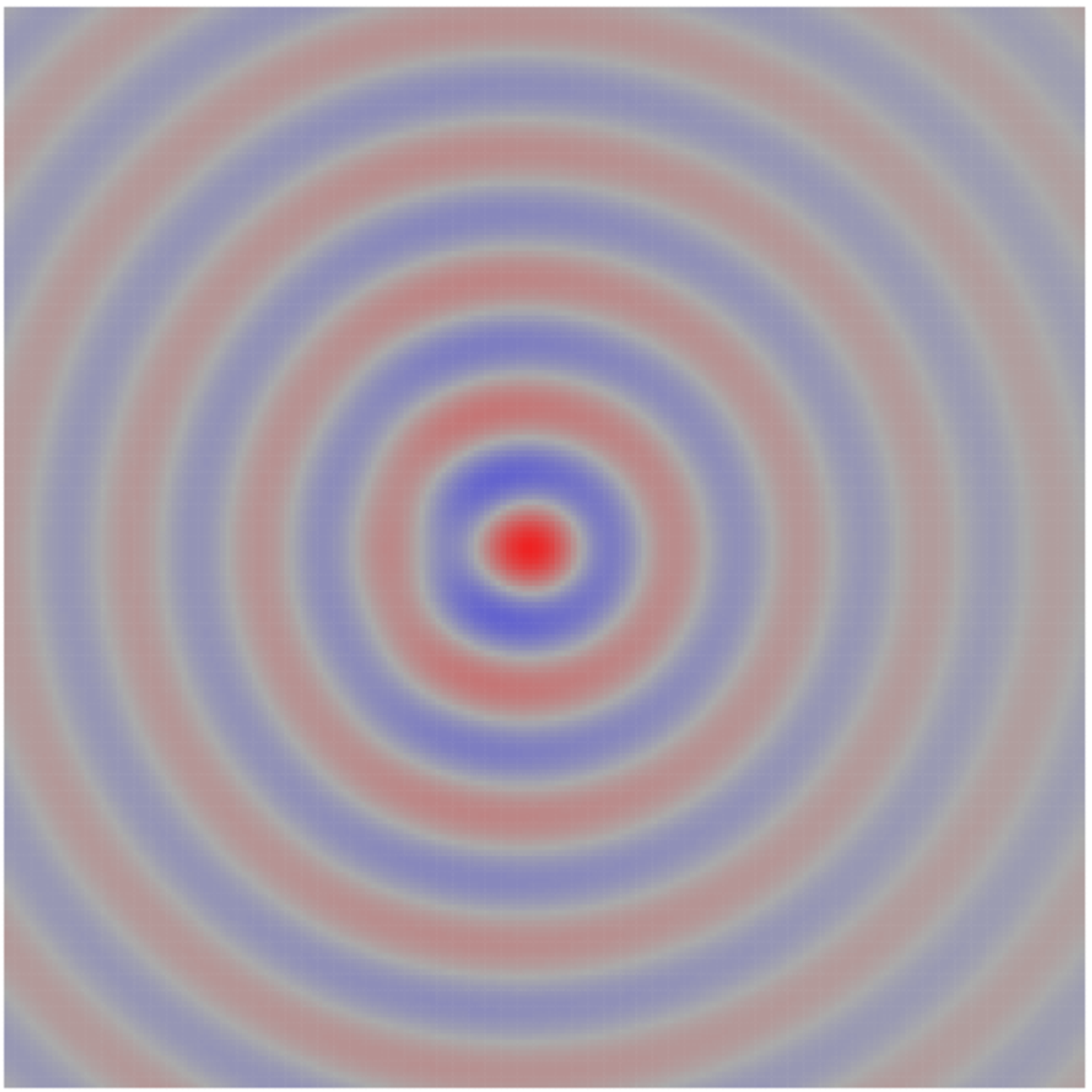} \includegraphics[width=0.23\textwidth]{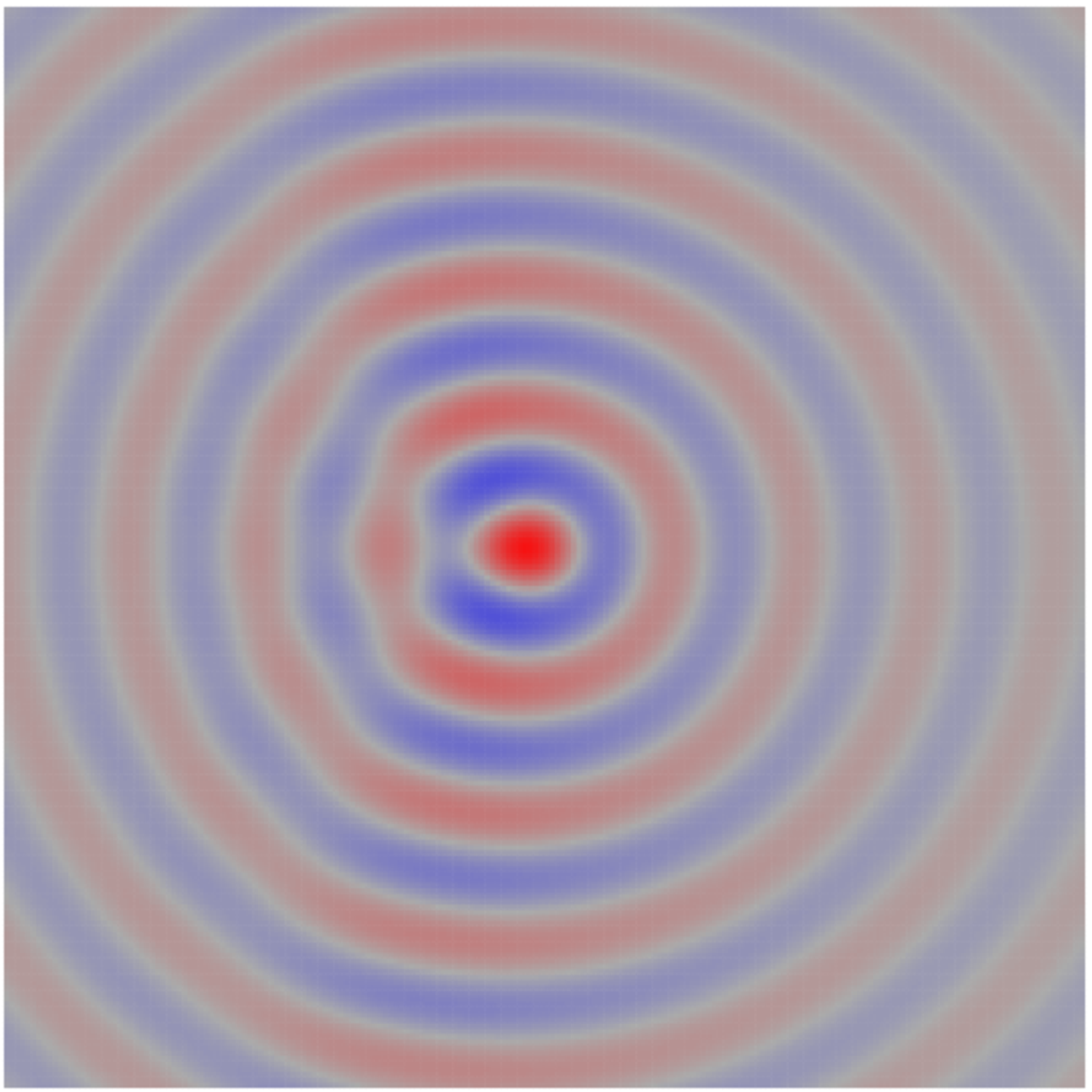} 
\includegraphics[width=0.23\textwidth]{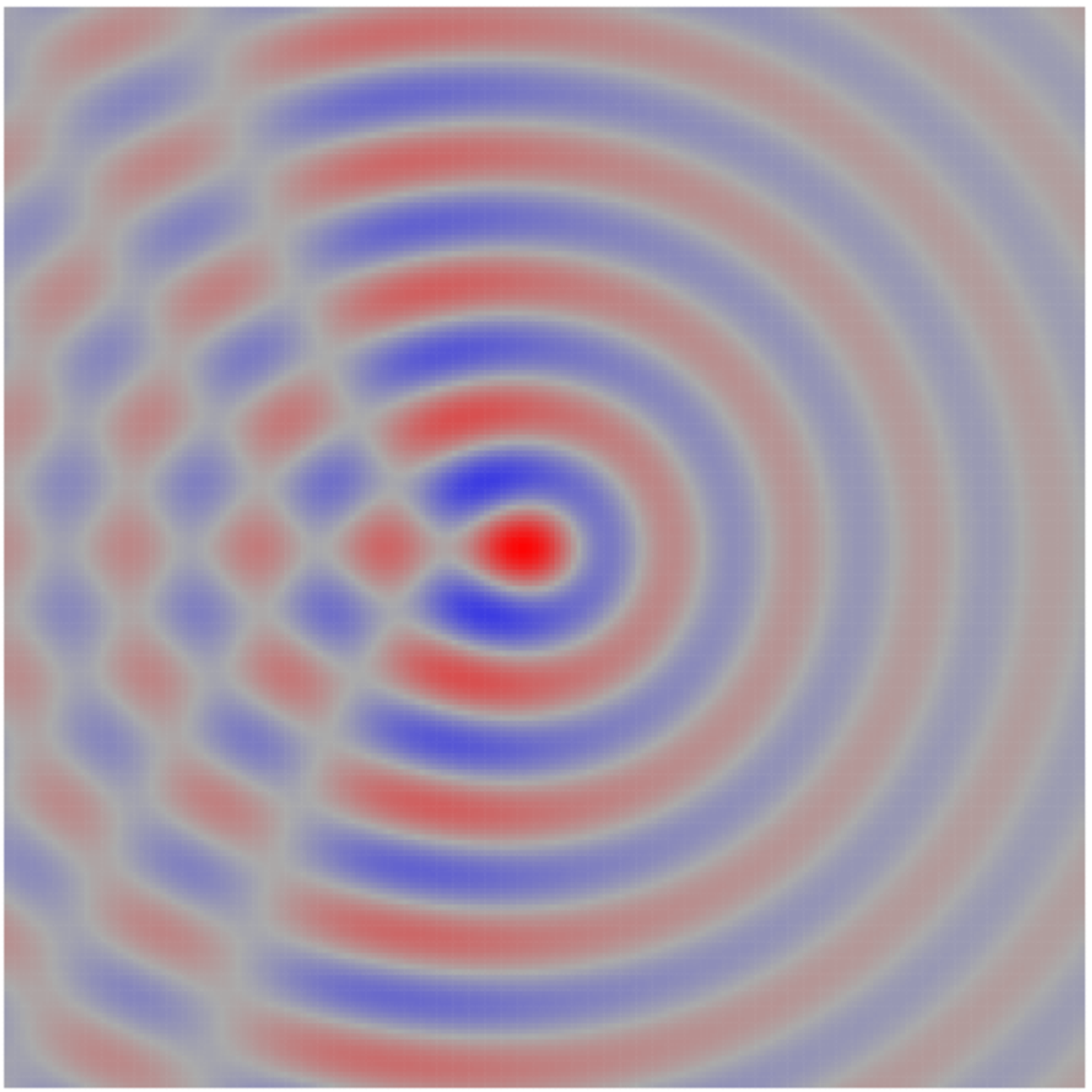} 
\includegraphics[width=0.23\textwidth]{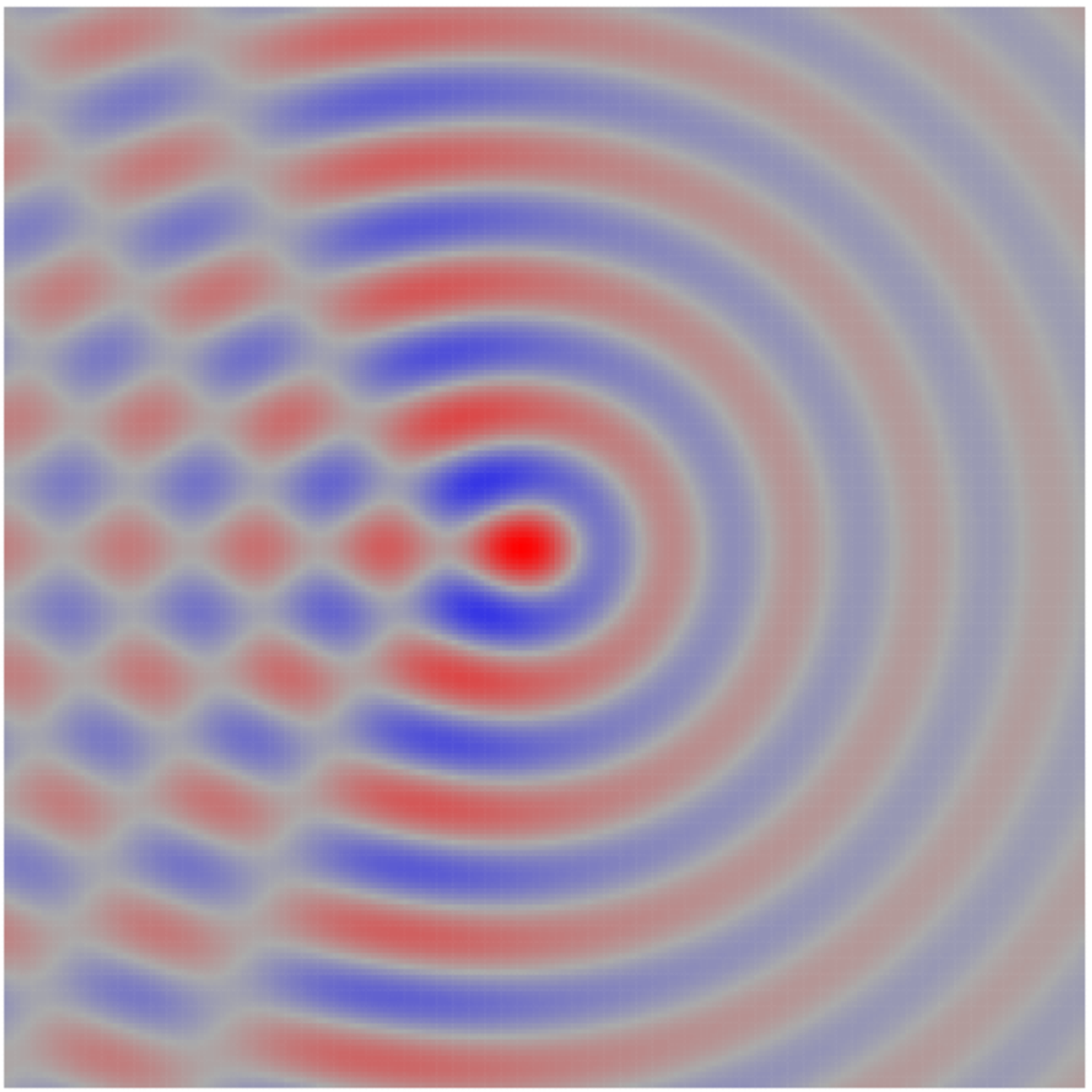} 
\vspace{8mm}

\textit{b.}\hspace{7mm}  \includegraphics[width=0.38\textwidth]{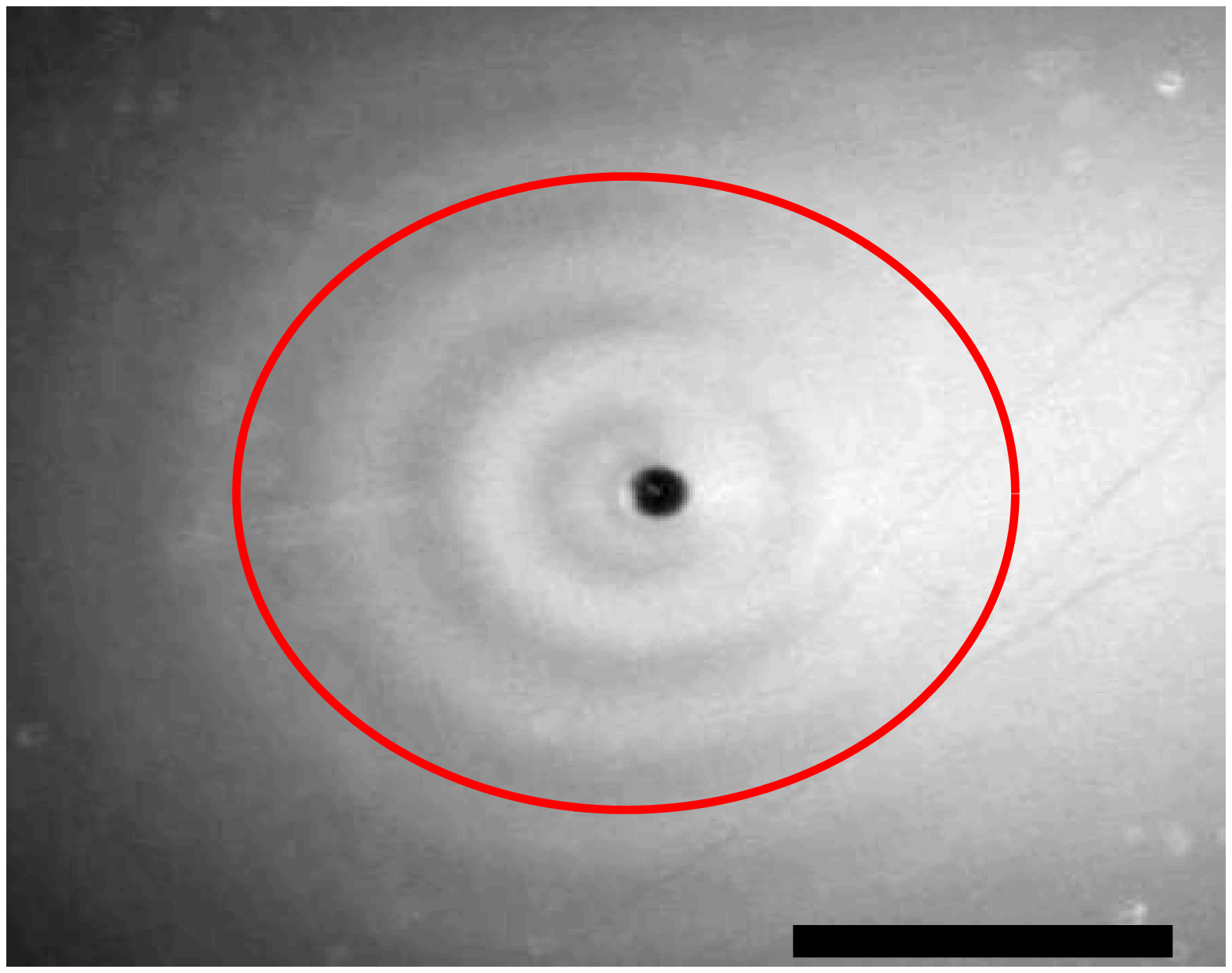}
\textit{c.}\hspace{5mm} \includegraphics[width=0.4\textwidth]{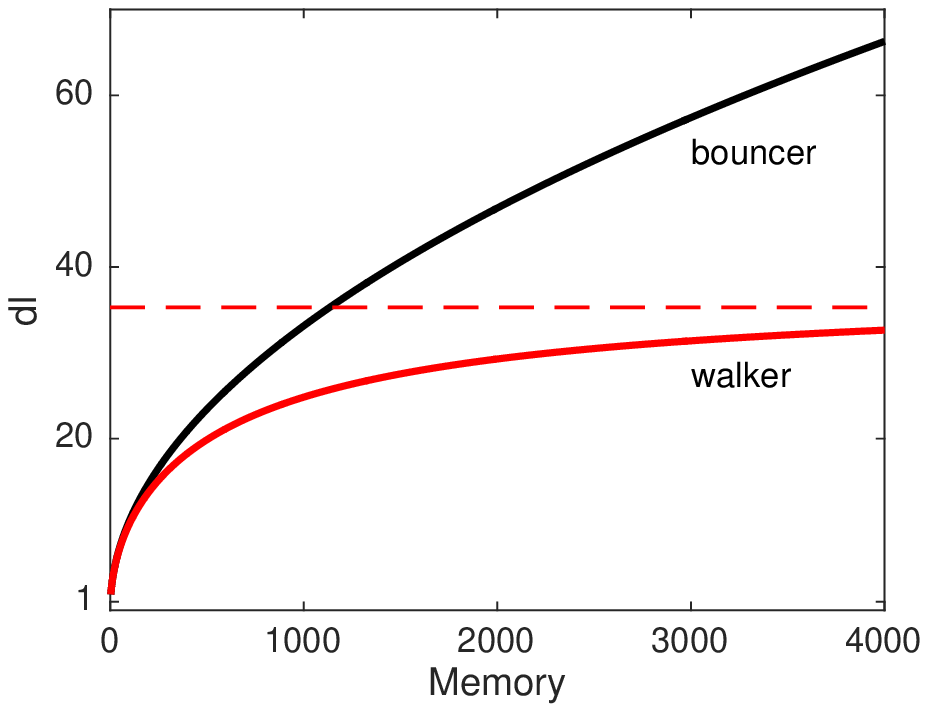} \hspace{5mm} 
\end{psfrags}
\caption{\textit{a.} Asymmetric wave field of a walker, calculated numerically from equation~\eqref{eq:walker}, for different values of the memory $\Me$. From left to right, $\Me=7.7$, $\Me=16$, $\Me=82$ and $\Me=447$ respectively. The walker velocity is fixed to $V_w = 10$ mm/s from left to right, and other parameters are given in table~\ref{tab:notations}. The field of view is $4 \times 4$~cm. \textit{b.} Anisotropic damping length $l(\theta)$, compared to the experimental wave-field of a walker. The red curve corresponds to equation \eqref{eq:damping_length_l} with $\mathcal{M}=10$ and other numerical values from table \ref{tab:notations}. In the experiment, the walking speed is 10 mm/s and $\mathcal{M}=10$. The scale bar is 1 cm. \textit{c.} Lateral damping length ($\theta=\pi/2$) for a bouncer (black curve, $V_w = 0$) and a walker (red curve, $V_w = 11$~mm/s), as a function of the distance to threshold $\mathcal{M}$.}
\label{fig:walker_waves}
\end{center}
\end{figure}

This sum is evaluated in Appendix~\ref{app:walker}. It yields
\begin{eqnarray}
\zeta_w(\mathbf{r},\tau) &=& \frac{B_F^+ k_F}{2\sqrt{D}} \cos\left(\tau+\theta_F^+ \right) \exp\left(- \frac{v r \cos\theta}{2D} \right) \label{eq:walker} \\
&\times& \int_{-\pi}^\pi d\varphi \frac{\exp[i k_F r \sin(\theta - \varphi)]}
{\sqrt{a- i k_F v \sin\varphi}} 
\exp{\left(-\frac{r}{\sqrt{D}}\sqrt{a - i k_F v \sin\varphi}\right)} \nonumber
\end{eqnarray}
where $\cos \theta = \mathbf{v} \cdot \mathbf{r} / (v r)$ and 
\begin{equation}
  a = \frac{1}{2\pi\Me} + \frac{v^2}{4D} \,.
\end{equation}
The corresponding wave-field can be calculated by numerically integrating equation~\eqref{eq:walker}. Some characteristic wave-fields are shown in figure~\ref{fig:walker_waves}a, for different values of the memory $\Me$. With increasing $\Me$, the sum includes significant contributions from an increasing number of past impacts. Some interferences appear in the wake of the walker, similarly to what is seen experimentally \citep{Eddi2011}. 

In the limit of small walking speed and low memory ($2\pi\Me k_F v\ll 1$), the integral \eqref{eq:walker} can be solved analytically (appendix~\ref{app:walker}):
\begin{equation}
\zeta_w(\mathbf{r},\tau) = \frac{B_F^+ k_F}{\sqrt{v^2 + \frac{2D}{\pi \Me}}} \cos\left(\tau+\theta_F^+ \right) J_0\left(k_F\left| \mathbf{r} + \frac{r}{\sqrt{v^2 + \frac{2D}{\pi \Me}}} \mathbf{v} \right|\right) \exp\left[- r / l(\theta) \right] \,,
\label{eq:walkerwave}
\end{equation}
where 
\begin{equation}
l(\theta) = \frac{2D}{v\cos\theta + \sqrt{ v^2 + \frac{2D}{\pi \Me}}}\,.
\label{eq:damping_length_l}
\end{equation}
Equation~\eqref{eq:walkerwave} corresponds to a Doppler-shifted Bessel profile which is exponentially attenuated on an length scale $l(\theta)$ that varies with direction $\theta$. The resulting wavelength is smaller in front of the walker and larger in its wake. The damping distance is smaller in front of the walker than in its wake, as illustrated in figure~\ref{fig:walker_waves}b. 

For either larger walking speed of larger memory, interferences between Bessel functions are more complex and solving the integral \eqref{eq:walker} is not straightforward. Nevertheless, an upper bound to the wave field can be found, as derived in appendix~\ref{app:walker}: 
\begin{equation}
\left| \frac{\zeta_w(\mathbf{r},\tau)}{\cos(\tau+\theta_F^+)}\right| \leq \frac{\pi B_F^+ k_F}{\sqrt{D a}} \exp\left[-\frac{r}{l(\theta)}\right] \,, \label{eq:limit}
\end{equation}
This bound indicates that the Faraday waves emitted by the walkers are always subjected to the same exponential spatial damping $l(\theta)$ defined in equation~\eqref{eq:damping_length_l}. 

In the case of a bouncer ($v=0$), this damping length $l(\theta) = \sqrt{2 \pi D \Me}$ is independent of $\theta$ (axisymmetric damping) and it grows indefinitely with increasing memory. At the Faraday threshold (infinite memory), a bouncer would generate waves that are not spatially damped. By contrast, waves are always spatially damped for a walker with finite speed $v$. Indeed, when $\Me \rightarrow \infty$, $l(\theta) \rightarrow 2 D / [v (1 + \cos \theta)]$. This horizontal asymptote is illustrated in figure~\ref{fig:walker_waves}c. Hence, equation \eqref{eq:limit} predicts a finite spatial damping even at infinite memory, and this for all directions except $\theta=\pi$.%The wave field of a walker is therefore attenuated in all directions (except in the backward direction $\theta = \pi$, but the estimation \eqref{eq:limit} at infinite memory is probably not meaningful at these positions). 

\subsection{Approximations specific to the experimental regime of walkers}
\label{sec:lowvisc}

The sub-threshold Faraday wave theory developed in previous sections is now particularised to the regime of walkers that corresponds to most experiments reported up to now \citep{Bush2015}. Typical values for each parameter are summarised in table \ref{tab:notations}. The liquid is silicone oil with density $\rho = 956\,\mathrm{kg}/\mathrm{m}^3$, surface tension $\sigma = 20.6 \times 10^{-3}\,\mathrm{N}/\mathrm{m}$, and viscosity $\nu = 20 \times 10^{-6}\,\mathrm{m}^2/\mathrm{s}$. The bath is vibrated sinusoidally at a frequency $\Omega/2\pi = 80$ Hz. 

In the limit of vanishing viscosity $\nu$, $\gamma_k \rightarrow 0$, and $f_k(i) \rightarrow \omega_k^2 -1$. Therefore, according to equation~\eqref{eq:th}, the instability threshold $\Gamma_F \rightarrow 0$ in $k = k_0$ such that $\omega_0^2 = \omega_{k_0}^2 = 1$. For a walker, $k_0 = 1319.5$ m$^{-1}$, which corresponds to a wavelength of 4.76~mm, already close to experimental observations \citep{Eddi2011}. The corresponding damping (at $\nu$ = 20~cS) is $\gamma_0 = \gamma_{k_0} = 0.277$. Although this value is not much smaller than 1, we can perform a Taylor expansion in $\gamma_0$ for each quantity of interest.

First, we prove in Appendix~\ref{app:low_viscosity} that the most unstable wave-number $k_F$, defined from equation~\eqref{eq:def_kf}, can be approximated by:
\begin{equation}
k_F= k_0 + \frac{1}{2c_0}\gamma_0^{3/2}-\frac{2+k_0c_0}{2k_0 c_0^2}\gamma_0^{2}+\frac{12+k_0 c_0}{8 k_0 c_0^2}\gamma_0^{5/2}+\left(\frac{1}{4k_0 c_0^2}-\frac{d_0}{8c_0^3}\right)\gamma_0^3+\mathcal{O}(\gamma_0^{7/2})
\label{eq:kf}
\end{equation}
where  $d_0=(\partial^2\omega_k^2/\partial k^2) (k_0)=24\sigma k_0/\rho\Omega^2$ and $c_0 = (\partial\omega_k/\partial k) (k_0)=(2(g+3\sigma k_0^2/rho )/\Omega^2)$ is the group velocity of free waves around the wave number $k_0$. For walkers, $c_0 = 0.97$~mm (which corresponds to a speed $\Omega c_0/2 = 0.24$~m/s) and the approximation \eqref{eq:kf} gives $k_F \simeq 1338.7$~m$^{-1}$ and $\lambda_f =  4.693$~mm, while a numerical computation of the exact solution of equation \eqref{eq:def_kf} without any approximation except the Floquet's first order truncation would give $k_F = 1334.8$ m$^{-1}$ and $\lambda_f=4.707$~mm. 

From Appendix~\ref{app:low_viscosity}, it follows that 
\begin{eqnarray}
\omega_F^2 &=& 1 + \gamma_0^{3/2} -\frac{2+k_0 c_0}{k_0 c_0}\gamma_0^2+ \frac{12+k_0 c_0}{4 k_0 c_0} \gamma_0^{5/2} +\mathcal{O}\left(\gamma_0^3\right) \nonumber \\
\gamma_F &=& \gamma_0+\frac{\gamma_0^{5/2}}{k_0 c_0}+\mathcal{O}\left(\gamma_0^3\right) \, .
\end{eqnarray}

Equation~\eqref{eq:th} for the corresponding instability threshold $\Gamma_{F}=\Gamma_k (k_F)$ is then approximated by:
\begin{equation}
\Gamma_F = \frac{\Omega^2}{g k_0}\gamma_0 - \frac{\Omega^2}{2 g k_0} \gamma_0^{3/2} + \mathcal{O}(\gamma_0^{5/2})
\label{eq:Gamma_F}
\end{equation}
The first term of $\Gamma_F$ corresponds to the threshold predicted from the damped Matthieu equation~\eqref{eq:DampedMatthieu} with a phenomenological factor $\alpha = 2$ \citep{Sampara2016}. Since for walkers $\gamma_0 = 0.277$, a limitation to this first term would yield $\Gamma_F = 5.41$. Considering instead the first two terms yields $\Gamma_F = 3.985$. The exact solution of equation~\eqref{eq:th} obtained numerically is $\Gamma_F = 4.144$, which is very close to experimental values. The correction in $\gamma_0^{3/2}$ is therefore a considerable improvement of this theory with respect to the damped Matthieu equation. 

The memory $\Me$ defined in equation~\eqref{eq:definition_memory} can be approximated by:
\begin{equation}
\Me = \frac{\mathcal{M}}{2\pi \gamma_0}\left( 1+ \frac{\gamma_0^{1/2}}{2}+\frac{\gamma_0}{4}+\mathcal{O}(\gamma_0^{3/2})\right)
\label{eq:definition_memoryApprox}
\end{equation}
It is therefore inversely proportional to the viscosity $\nu$ and an inviscid fluid would yield a theoretically infinite memory. In the case of walkers, $(1+\gamma_0^{1/2}/2+\gamma_0/4)/2\pi\gamma_0 \sim 0.76$, which is fairly close to unity. In previous work on walkers, $\Me$ has often been mixed up with the distance to threshold $\mathcal{M}$, without much consequence. Close to threshold ($\Gamma \lesssim \Gamma_F$), the damping factor of short-lived Faraday waves $\delta_F^- = -2\gamma_0+\mathcal{O}(\gamma_0^{3/2})$ is significantly more negative than the damping of long-lived waves $\delta_F^+ = - 1/ (2 \pi \Me)$. 

The diffusion coefficient $D$ associated to long-lived Faraday waves and defined in equation~\eqref{eq:D_exact} is approximated by 
\begin{equation}
D = \frac{c_0^2}{2\gamma_0}\left(1+\frac{\gamma_0^{1/2}}{2}+\mathcal{O}(\gamma_0)\right)
\label{eq:D_approx}
\end{equation}
which is about $D \simeq 4.40$ $mm^2$ for a walker. 

The phase shift of long-lived and short lived Faraday waves near the Faraday threshold is:
\begin{equation}
\theta_F^{\pm} = \pm \pi/4 +\mathcal{O}(\gamma_0)\,.
\end{equation}
and consequently, 
\begin{equation}
A_F^{\pm} = v_F \cos \left(\tau_i \mp \frac{\pi}{4} \right)  e^{-\delta_F^{\pm} \tau_i} + \mathcal{O}(\gamma_0) \mbox{ and } B_F^+ = v_F \cos \left( \tau_i - \frac{\pi}{4} \right) + \mathcal{O}(\gamma_0) 
\end{equation}

The long-lived impulse response \eqref{eq:zeta2} then becomes:
\begin{eqnarray}
  \zeta_1(r,\tau) & \simeq & v_F k_F 
  \sqrt{\frac{\pi}{D (\tau - \tau_i)}} J_0(k_F r) 
  \cos\left(\tau_i -  \frac{\pi}{4}\right) 
  \cos\left(\tau + \frac{\pi}{4}\right) \nonumber \\ 
  && \times \exp\left[-\frac{\tau-\tau_i}{2\pi \Me} - 
    \frac{r^2}{4 D (\tau-\tau_i)} \right]\, , \label{eq:Faradaywave}
\end{eqnarray} 
which reads in dimensional form,
\begin{eqnarray}
  \zeta_1(r,t) & \simeq & v_F k_F
  \sqrt{\frac{2\pi}{D\Omega (t - t_i)}} J_0(k_F r) 
  \cos\left(\frac{\Omega t_i}{2} -  \frac{\pi}{4}\right) 
  \cos\left(\frac{\Omega t}{2} + \frac{\pi}{4}\right) \nonumber \\ 
  && \times \exp\left[-\frac{(t-t_i)}{\Me T_F} - 
    \frac{r^2}{2 D\Omega (t - t_i)} \right] \label{eq:Faradaywave_dimensional}
\end{eqnarray}

The upper bound (equation~\eqref{eq:limit}) on the wave-field of a walker that moves at constant normalised speed $v$ becomes
\begin{equation}
\left| \frac{\zeta_w(\mathbf{r},\tau)}{\cos (\tau + \pi/4)} \right| \leq \frac{2 \pi v_F k_F \cos (\tau_i - \pi/4)}{\sqrt{v^2 + \frac{c_0^2}{\mathcal{M}}}} \exp \left[ - \frac{r}{l(\theta)} \right]
\end{equation}
with 
\begin{equation}
l(\theta) \simeq \frac{c_0^2}{2 \gamma_0 \left[ v \cos \theta + \sqrt{v^2 + c_0^2/\mathcal{M}} \right]}
\end{equation}
For distance to threshold $\mathcal{M} \gg c_0^2 / v^2$ (which is about 490 for walkers),  
\begin{equation}
l(0) \simeq \frac{c_0^2}{4 \gamma_0 v} \simeq 1.9~\textrm{cm}, \mbox{ and } l(\pi) \simeq \frac{\mathcal{M}}{\gamma_0}
\end{equation}
The damping length $l$ can therefore be as large as wanted in the wake of the walker ($\theta = \pi$), but it is strongly limited in front of the walker, independently of $\mathcal{M}$. Nevertheless, $l(0)$ could be increased by decreasing $\gamma_0$, i.e. by decreasing viscosity and associated dissipation.

At sufficiently low $\mathcal{M}$, given by equation~\eqref{eq:walkerwave} is approximated by:
\begin{equation} 
\zeta_w(\mathbf{r},t) \simeq \frac{v_F k_F}{\sqrt{v^2 + \frac{c_0^2}{\mathcal{M}}}} \cos \left( \tau_i - \frac{\pi}{4} \right) \cos \left( \tau + \frac{\pi}{4} \right) J_0 \left[ k_F \left| \mathbf{r} + \frac{r}{\sqrt{v^2 + \frac{c_0^2}{\mathcal{M}} } } \mathbf{v} \right| \right] e^{-r / l(\theta)}
\label{eq:walkerwaveApprox}
\end{equation}
This approximation is only valid when $k_F v / a \ll 1$ (appendix~\ref{app:walker}), which here reduces to
\begin{equation}
\mathcal{M} \ll \left(\frac{k_F v}{\gamma_0} - \frac{v^2}{c_0^2}\right)^{-1}
\end{equation}
This limit is $\mathcal{M} \ll 4.9$ for a walker at speed $V_w = 11$~mm/s. Furthermore, since $\mathcal{M} \ll \frac{c_0^2}{v^2} \simeq 490$, the Doppler shift along the walker trajectory in equation~\eqref{eq:walkerwaveApprox} can be approximated by
\begin{equation}
k_\mathrm{Doppler} = k_F \pm \frac{v}{c_0} \sqrt{\mathcal{M}}\, .
\end{equation}
This expression is very similar to the one obtained numerically in \citet{Milewski2015}. %\citet{Milewski2015} computed numerically that the Doppler effect shifts the wavelength as $k_\mathrm{Doppler} = k_F (1 \pm v k_F/\omega_F)$ for a distance to threshold $\sim 7$ leading to a factor $\sim 2$ difference with our prediction.

\section{Experimental observation of the surface waves emitted by a localised impact}

\label{sec:Experiments}

%Since the seminal works of \citep{Couder2005, Protiere2006}, it is well accepted that the walkers are driven by the waves generated by their previous rebounds. The space and time dependencies of the surface waves are crucial to predict the behaviour of a walker. Up to now, the wavefield generated by walkers has been described at the scale of one Faraday period \citep{Eddi2011,Molacek2013b}. Those descriptions have proven to model efficiently the regular bouncing of a single walker in various configurations.
%The interaction of several walkers is more complex and leads to possible destabilization of the regular bouncing that drives a single walker. An accurate space and time description of the surface waves is thus essential in this context.

This section aims at validating experimentally the theory of sub-threshold surface waves developed in section~\ref{sec:theory}. Specifically, we characterise (i) the spatial damping of the Faraday waves upon a single impact, (ii) the phase shift between these waves and the shaker position, and (iii) the dependence of the wave amplitude on the phase of impact. These three points are major additions of the present theory with respect to previous models of the Faraday waves generated by walkers \citep{Eddi2011, Molacek2013b}. 

%since the present theory predicts an exponential diffusive behaviour whereas other theories describe a direct exponential damping . %This parameter, although crucial to describe the surface waves is often not considered 

\subsection{Experimental set-up}

\begin{figure}
\begin{psfrags}
\psfrag{S}[c][c]{Shaker}\psfrag{Tv}[l][l]{}\psfrag{Tpv}[l][l]{}\psfrag{Gb}[l][l]{Glass bead}
\psfrag{Ob}[l][l]{Oil reservoir}\psfrag{A}[l][l]{$g(t)$}\psfrag{Syn}[l][l]{Synthetic}\psfrag{Sch}[l][l]{Schlieren}
\psfrag{sub}[l][l]{Shallow region}\psfrag{acc}[l][l]{Accelerometer}\psfrag{Vj}[r][r]{Vice jaw}
\includegraphics[width=0.9\textwidth]{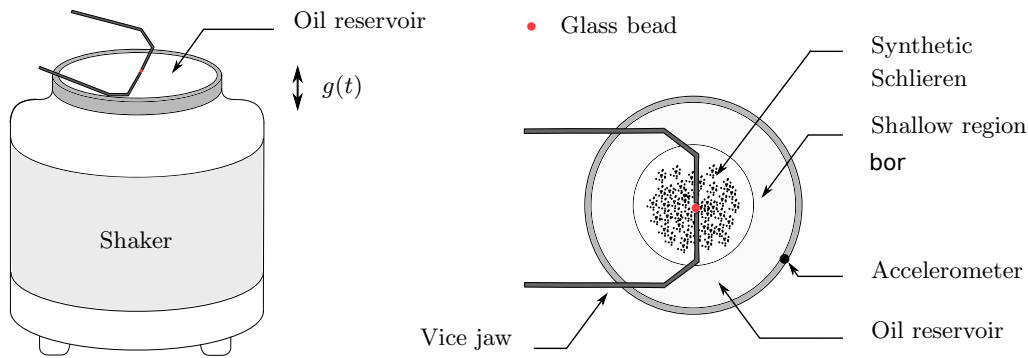}
\end{psfrags}
\caption{Tilted view (left) and top view (right) of the experimental set-up. A large transparent reservoir of oil is mounted on top of a shaker. Below the reservoir a transparent sheet is printed with a synthetic Schlieren. The border of the reservoir is a shallow region where surface waves are strongly damped. A glass bead is released on demand above the reservoir with a vice jaw opened with a translation stage. The acceleration of the reservoir is measured with an accelerometer placed on the border of the reservoir.}
\label{fig:experimental_set-up}
\end{figure}

The goal of the experiment is to finely quantify the spatio-temporal evolution of the surface waves generated by the single localised impact of a bead. The experimental set-up, depicted in figure \ref{fig:experimental_set-up}, comprises a transparent cylindrical reservoir of diameter 22~cm filled with about 5~mm of silicone oil (Sigma-Aldrich). The reservoir is delimited by a shallow region (width 3~cm, depth $< 1$~mm), thanks to which surface waves are strongly damped. This trick allows to eliminate the additional waves generated by the meniscus at the boundaries of the reservoir \citep{Bechhoefer1995, NguyemThuLam2011}. The reservoir is placed on top of an electromagnetic shaker (V400LT, Data Physics). The shaker motion is accurately controlled via an accelerometer tightly fixed on the reservoir, a computer interface and a feedback loop, as detailed in \citet{Sampara2016}. The resulting effective gravity acceleration is $g(t) = g (1+\Gamma\cos \Omega t)$. The frequency is fixed to $\Omega= 2\pi\,80$ s$^{-1}$. 

The bead is made of glass and its diameter is 1~mm. A device that can release the bead on demand is placed at 1~cm above the bath. It consists in a vice with two jaws that grip the bead. It is mechanically insulated from the shaker, so it is not sensitive to its vibrations. A translation stage allows to open the vice and release the bead. The bead does not acquire any significant velocity from the vice, so it starts falling from rest. 

Measurements must be gathered over many Faraday periods, with a time resolution well below the shaker period $2\pi / \Omega$. Therefore, we here use high-speed imaging (Phantom MIRO110), with an acquisition frequency of 4000 frames per second (i.e. 50 frames per period of the shaker). The shaker is illuminated with 4 home-made high-power LEDs to ensure strong illumination of the waves while limiting the associated heating. Thanks to a semi-refringent mirror placed at 45$^o$ above the shaker, optical paths of both illumination and recording are vertical near the bath surface. Thanks to a macro lens \textit{Zeiss Milvus 2/100M}, the spatial resolution is brought down to 70 $\mu$m by pixel. The camera is synchronised with the shaker thanks to a DACQ (National Instruments) and a \textit{Labview} interface that returns the impact phase $\tau_i$, with respect to the Faraday period $T_F$. Details on this crucial synchronisation are provided in Appendix \ref{app:experimental_set_up}.

The free-surface synthetic Schlieren technique developed by \citet{Moisy2009} is well-adapted to measure wave elevations in real time. It consists in placing a transparent sheet printed with randomly spaced dots (dot diameter 100 $\mu$m, 50\% density) over a white background at the bottom of the reservoir. The image of these dots moves owing to the light refraction at the free surface. The displacement of the imaged dots is therefore related to the local slope of the free surface. This technique was already successfully applied to measure the 3D shape of the long-lived Faraday waves generated by walkers, with an acquisition frequency lower than $\Omega / (2\pi)$ \citep{Eddi2011, Damiano2016}. However, in its original form, it is not well-suited to the measurements intended here. Indeed, it requires a large field of view that must not be obstructed by the device which would drop the bead. Moreover, owing to the limited amount of information (3 GB) that can be acquired by a high-speed camera in a single shot, the recording duration at full spatial resolution would be a fraction of a second, which is definitely not sufficient to capture the long-term behaviour of Faraday waves. Finally, the reconstruction algorithm of Moisy's technique would demand large computational resources. In order to circumvent these limitations, we here use an adaptation of Moisy's technique. Since the wave field is known to be axisymmetric, we restrict the field of view to a radial zone of $2\times15$~mm from the impact point. This allows to record much longer films (several seconds at 4000~fps), without any obstruction of the field of view by the bead-releasing device. The displacement of the imaged dots is calculated along a radius from impact point. The one-dimensional radial gradient of the elevation $\zeta(r,t)$ is then calculated with a 1D adaptation of the reconstruction algorithm detailed in \citet{Moisy2009}. The restriction to a quasi-1D wave pattern severely decreases the computational time of the reconstruction. However, it can be shown that the 1D algorithm does not benefit anymore from the error compensation mechanism present in the 2D algorithm. In order to limit the measurement error, we first remove most of the noise with deshaking and median filters. We also avoid the error build-up inherent to the spatial integration of $\partial_r \zeta$ to obtain $\zeta(r,t)$. We instead directly compare the measured $\partial_r \zeta$ to the theoretical prediction 
\begin{eqnarray}
\frac{\partial \zeta_1}{\partial r}(r,t) &=&-v_F k_F
  \sqrt{\frac{2\pi}{D\Omega (t - t_i)}} 
  \cos\left(\frac{\Omega t_i}{2} -  \frac{\pi}{4}\right) 
  \cos\left(\frac{\Omega t}{2} + \frac{\pi}{4}\right) \nonumber \\ 
  &\times &  \left[J_1(k_F r) +\frac{r\,J_0(k_F r)}{D\Omega (t - t_i)}\right]\exp\left[-\frac{(t-t_i)}{\Me T_F}- 
    \frac{r^2}{2 D\Omega (t - t_i)}\right]\, .
    \label{eq:wave_gradient}
\end{eqnarray}

Several input parameters are systematically varied, including the impact phase of the bead relative to the shaker, the shaker amplitude and corresponding distance to threshold $\mathcal{M}$, and the liquid viscosity (5~cS and 20~cS silicone oil). %3 impacts are recorded for each configuration of parameters. 

\subsection{Spatio-temporal evolution of the wave field}

\begin{figure}
\begin{center}
\begin{psfrags}
\psfrag{0}[c][c]{0}\psfrag{0.1}[c][c]{0.1}\psfrag{0.2}[c][c]{0.2}\psfrag{0.3}[c][c]{0.3}\psfrag{0.4}[c][c]{0.4}
\psfrag{0.5}[c][c]{0.5}\psfrag{0.6}[c][c]{0.6}\psfrag{0.005}[c][c]{5}\psfrag{0.01}[c][c]{10}\psfrag{0.015}[c][c]{15}
\psfrag{a}[ct][cb]{a}\psfrag{b}[ct][cb]{b}\psfrag{time}[cb][ct]{$t$ (s)}\psfrag{distance}[c][c]{$r$ (mm)}\psfrag{Wavegradient}[c][c]{Wave gradient}\psfrag{spatiotemp1}[cb][ct]{}\psfrag{spatiotemp2}[cb][ct]{}
\psfrag{m1}[l][l]{$-10^{-3}$}\psfrag{m2}[l][l]{$-5\,10^{-4}$}\psfrag{m3}[l][l]{0}\psfrag{m4}[l][l]{$5\,10^{-4}$}\psfrag{m5}[l][l]{$10^{-3}$}
\includegraphics[width=0.9\textwidth ]{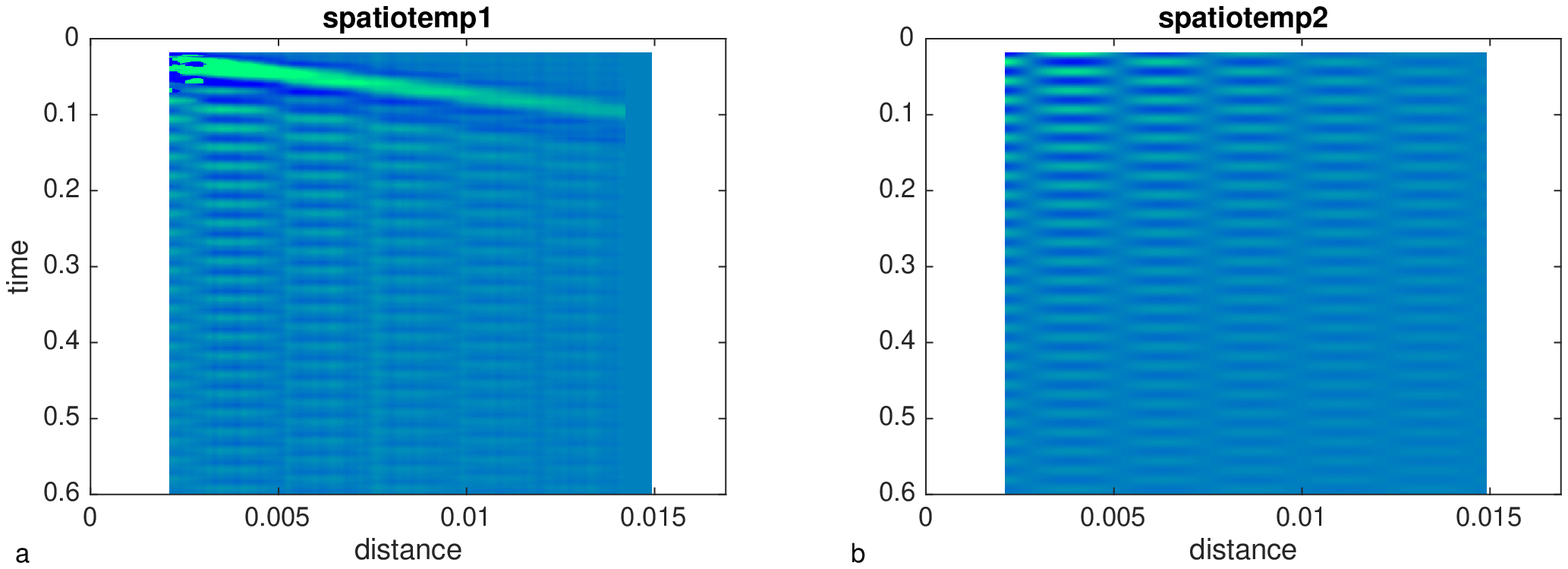}\hspace{4mm}\includegraphics[width=0.05\textwidth, height= 4.5cm]{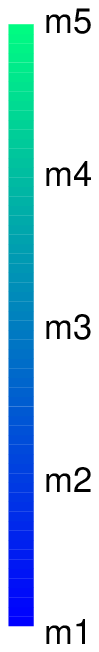}
\end{psfrags}
\caption{Qualitative comparison of the spatio-temporal evolution of the radial wave gradient $\partial_r \zeta (r,t)$. The impact is located in $(r,t) = (0,0)$. \textit{a.} Experiment at $\mathcal{M} = 30$ and $\nu=20$ cS. They are gathered in table \ref{tab:notations}. \textit{b.} Theoretical prediction \eqref{eq:wave_gradient}.
}
\label{fig:wavegradient_comparison}
\end{center}
\end{figure}

A typical evolution of the wave field is represented in the space-time diagram of figure~\ref{fig:wavegradient_comparison}a. We observe that the impact first generates a large capillary wave packet that travels radially outwards with an almost constant velocity of 16~cm/s. The long-lived Faraday waves described in section~\ref{sec:theory} are observed in the wake of this wave packet. They have a pulsation $\Omega/2$ and a wavelength $2\pi / k_F \simeq 4.7$~mm. The surface appears still in the region that is not yet visited by the wave packet (i.e. in the upper-right corner of figure~\ref{fig:wavegradient_comparison}a). 

This experiment can be qualitatively compared to the theoretical prediction \eqref{eq:wave_gradient}, where the amplitude of this latter is fitted on the experiments (figure~\ref{fig:wavegradient_comparison}b). Overall, this comparison is promising. However, the capillary wave packet is not present in the prediction, which is expected since only long standing Faraday waves were considered in equation~\eqref{eq:wave_gradient}. This latter also predicts waves of finite amplitude everywhere on the bath at $t > t_i$, independently of the passage of the capillary wave packet, which was hard to observe experimentally. Nevertheless, their amplitude remains very small shortly after impact and far from the impact point, which may be the reason why the bath appears still this region of the experimental spatio-temporal diagram. 

\begin{figure}
\begin{center}
\begin{psfrags}
\psfrag{0}[c][c]{0}\psfrag{1}[c][c]{1}\psfrag{2}[c][c]{2}\psfrag{3}[c][c]{3}\psfrag{4}[c][c]{4}
\psfrag{-1}[c][c]{-1}\psfrag{-2}[c][c]{-2}\psfrag{-3}[c][c]{-3}\psfrag{-4}[c][c][0.7]{-4}
\psfrag{10}[c][c][0.9]{10}\psfrag{30}[c][c]{}\psfrag{50}[c][c]{}\psfrag{70}[c][c]{}
\psfrag{20}[c][c]{20}\psfrag{40}[c][c]{40}\psfrag{60}[c][c]{60}
\psfrag{0.5}[c][c]{}\psfrag{1.5}[c][c]{}\psfrag{2.5}[c][c]{}\psfrag{3.5}[c][c]{}
\psfrag{a}[ct][cb]{a}\psfrag{b}[ct][cb]{b}
\psfrag{am}[cr][cr][0.7]{$-10^{-4}$}\psfrag{a0}[cr][cr][0.7]{0}\psfrag{ap}[cr][cr][0.7]{$10^{-4}$}
\psfrag{Rwg2}[cb][ct][0.8]{wave grad.}\psfrag{Ndt2}[cb][ct][0.8]{$\tau/\pi$}
\psfrag{Rwg}[cb][ct]{Radial wave gradient}\psfrag{Ndt}[ct][cb]{Normalized time, $\tau/\pi$}\psfrag{Ndr}[ct][cb]{Normalized radius, $r/\lambda_F$}\psfrag{Temporal}[c][c]{}\psfrag{Spatial}[c][c]{}
\includegraphics[width=\textwidth]{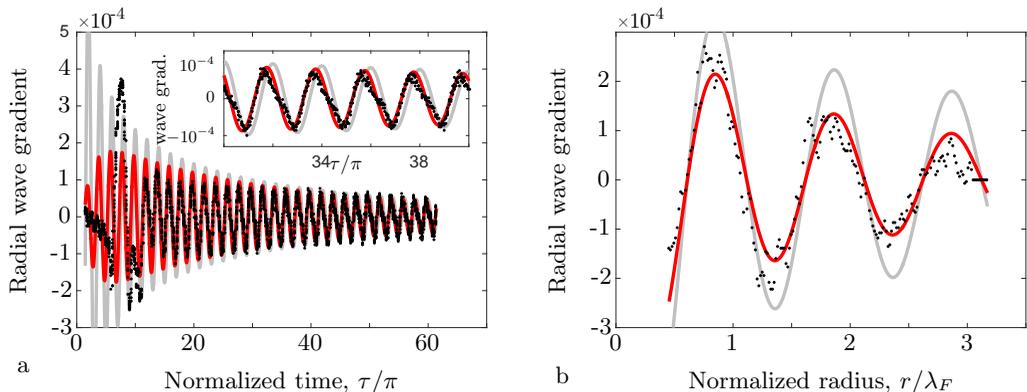}
\end{psfrags}
\caption{Radial gradient of surface waves, \textit{a.} versus time at a fixed position, and \textit{b.} versus position at a fixed time. The experimental data are a subset of those represented in figure~\ref{fig:wavegradient_comparison}a. Black dots correspond to the gradient measured experimentally. The grey and red solid curves correspond to equations~\eqref{eq:OldWaveField1impact} and \eqref{eq:wave_gradient}, respectively. The inset in \textit{a.} is a zoom of the main graph on a smaller time window. }
\label{fig:wavegradient_spatial_time_evolution}
\end{center}
\end{figure}

A more quantitative comparison is provided in figures~\ref{fig:wavegradient_spatial_time_evolution}a~and~\ref{fig:wavegradient_spatial_time_evolution}b, where the radial gradient of the wave is represented as a function of time (at a given position), and as a function of space (at a given time), respectively. The theoretical prediction \eqref{eq:wave_gradient} of long-lived Faraday waves is compared to both experiments and the previous model of  \citet{Molacek2013b} (gradient of equation~\ref{eq:OldWaveField1impact}). The amplitude factors ($v_F$ in equation~\eqref{eq:wave_gradient}, and $A \sin \Phi$ in equation~\eqref{eq:OldWaveField1impact}) are fitted on experimental data in the region $(r,t)$ that corresponds to after the passage of the capillary wave packet. 

In figure~\ref{fig:wavegradient_spatial_time_evolution}a, it is seen that both models capture well the long-term evolution of the wave at a given position. Nevertheless, a zoom on a small time window (inset of figure~\ref{fig:wavegradient_spatial_time_evolution}a) reveals that the factor $\cos(\Omega t / 2 +\pi/4)$ of equation~\eqref{eq:wave_gradient} fits experimental data better than the factor $\cos(\Omega t/2)$ of equation~\eqref{eq:OldWaveField1impact}. In the short-term (i.e. before the passage of the capillary wave packet), the experimental wave amplitude is very small. None of the models does accurately capture this behaviour, although the addition of short-lived waves in equation~\eqref{eq:wave_gradient} might improve the prediction. However, equation~\eqref{eq:OldWaveField1impact} diverges at short times, while equation~\eqref{eq:wave_gradient} grows from zero according to an exponential-diffusive behaviour, which is much more realistic. 
Moreover in equation \eqref{eq:OldWaveField1impact}, the information of impact is immediately transmitted everywhere on the bath. By contrast, the present theory predicts that after an impact the bath is first quiet and destabilizes progressively owing to diffusive spreading.

The radial evolution of the wave at a given time, seen in figure~\ref{fig:wavegradient_spatial_time_evolution}b, is qualitatively similar for both models and experiment. Nevertheless, the spatial damping predicted by equation~\eqref{eq:wave_gradient} is closer to experiments than the one of equation~\eqref{eq:OldWaveField1impact}. 

\begin{figure}
\begin{center}
\begin{psfrags}
\psfrag{0}[c][c][0.7]{0}%\psfrag{1}[c][c][0.7]{1}\psfrag{2}[c][c][0.7]{2}\psfrag{3}[c][c]{3}\psfrag{4}[c][c]{4}
%\psfrag{-1}[c][c][0.7]{-1}\psfrag{-2}[c][c]{-2}\psfrag{-3}[c][c]{-3}\psfrag{-4}[c][c][0.7]{-4}
%\psfrag{10}[c][c][0.9]{10}
\psfrag{1}[r][r]{$1$}\psfrag{0.1}[r][r][1]{$0.1$}\psfrag{10}[c][c]{10$^1$}\psfrag{102}[c][c]{10$^2$}
\psfrag{0}[c][c][0.7]{0}\psfrag{50}[c][c][0.7]{50}\psfrag{25}[c][c][0.7]{25}\psfrag{100}[c][c][0.7]{100}\psfrag{150}[c][c][0.7]{150}
\psfrag{20}[c][c][0.7]{20}\psfrag{40}[c][c][0.7]{40}\psfrag{60}[c][c][0.7]{60}
\psfrag{aa}[l][l][0.8]{20 cst Oil}
\psfrag{bb}[l][l][0.8]{$\Me = \mathcal{M}\Me^*$ [20 cst]}
\psfrag{cc}[l][l][0.8]{5 cst Oil}
\psfrag{dd}[l][l][0.8]{$\Me = \mathcal{M}\Me^*$ [5 cst]}
\psfrag{Eddi}[l][l][0.8]{$\Me = \mathcal{M}$}
\psfrag{dtc}[cb][ct]{Decay time constant, $\Me T_F$ (s)}\psfrag{Memory}[ct][cb]{Distance to threshold, $\mathcal M$}\includegraphics[width=0.8\textwidth]{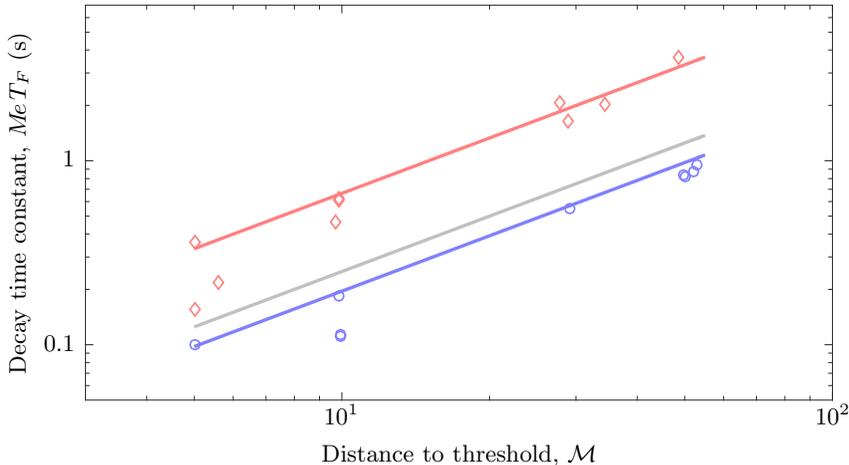}
\end{psfrags}
\caption{Color online. Temporal damping $\Me T_F$, as a function of the distance to threshold $\mathcal{M}$. Red diamonds (\textcolor{light-red}{$\lozenge$})and blue circles (\textcolor{light-blue}{$\circ$}) correspond to experiments with silicone oil of viscosity 5~cS and 20~cS, respectively. Red and blue solid lines correspond to equation~\eqref{eq:definition_memoryApprox} with the corresponding viscosity. The grey line represents $\Me T_F$ = $\mathcal M$.}
\label{fig:decay_time_constant}
\end{center}
\end{figure}

The temporal damping of Faraday waves (envelope of the curve in figure~\ref{fig:wavegradient_spatial_time_evolution}a) was already observed by \citet{Eddi2011}. It was then interpreted as the memory $\Me$ of the waves, which revealed to be a crucial parameter for the dynamics of walkers \citep{Bush2015}. However, the definition given by \citet{Eddi2011} differs from equation~\eqref{eq:definition_memoryApprox}, as the former is independent of viscosity while the latter is inversely proportional to viscosity in first approximation. In order to validate equation~\eqref{eq:definition_memory}, we have measured the damping time $\Me T_F$ as a function of the distance to threshold $\mathcal{M}$ for two different silicone oils, of kinematic viscosity 5~cS and 20~cS respectively (figure~\ref{fig:decay_time_constant}). The comparison with the theoretical prediction~\eqref{eq:definition_memoryApprox} is excellent. In particular $\Me T_F$ is proportional to $\mathcal{M}$ and inversely proportional to viscosity. 

\subsection{Wave amplitude versus impact phase}

%Another parameter of importance for the created waves is the phase of impact defined by
%\begin{equation}
%\phi = 2\pi (t_i[T_F])/T_F\,.
%\end{equation}  

\begin{figure}
\begin{center}
\begin{psfrags}
\psfrag{0}[c][c]{0}\psfrag{1}[c][c]{1}\psfrag{-1}[c][c]{-1}\psfrag{0.5}[c][c]{0.5}\psfrag{-0.5}[c][c]{-0.5}
\psfrag{10}[c][c][0.9]{10}\psfrag{-4}[c][c][0.7]{-4}
\psfrag{30}[c][c]{30}\psfrag{32}[c][c]{32}\psfrag{34}[c][c]{34}\psfrag{36}[c][c]{36}\psfrag{38}[c][c]{38}\psfrag{40}[c][c]{40}
\psfrag{p}[c][c]{$\pi/2$}\psfrag{2p}[c][c]{$\pi$}\psfrag{3p}[c][c]{$3\pi/2$}\psfrag{4p}[c][c]{$2\pi$}
\psfrag{a}[ct][cb]{a}\psfrag{b}[ct][cb]{b}
\psfrag{time}[c][c]{time (s)}\psfrag{distance}[c][c]{Distance from impact (mm)}\psfrag{Rwg}[cb][ct]{Radial wave gradient}\psfrag{Ndt}[ct][cb]{Normalized time $\tau/\pi$}\psfrag{Ndr}[c][c]{Normalized radius $r/\lambda_f$}\psfrag{Temporal}[c][c]{}\psfrag{Spatial}[c][c]{}\psfrag{ip}[ct][cb]{Impact phase $\tau_i$}\psfrag{Nafwg}[cb][ct]{Normalized amplitude}
\includegraphics[width=0.5\textwidth]{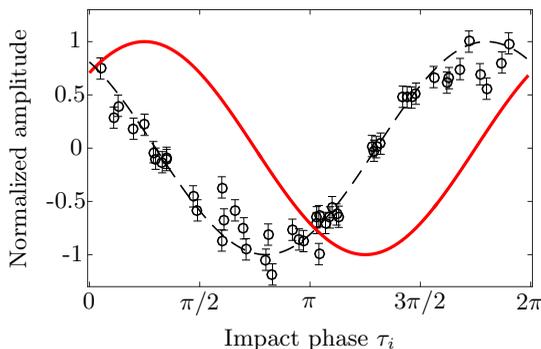}
\end{psfrags}
\caption{Normalised amplitude of the Faraday waves (at fixed position $r$ and time $t$), as a function of impact phase $\tau_i$. Symbols correspond to experimental data with oil viscosity $\nu=$20 cS, $\Omega= 80\times 2\pi$ s$^{-1}$ and $\mathcal{M}=30$. The red curve is the theoretical prediction $\cos(\tau_i - \pi/4)$ from the impulse response \eqref{eq:wave_gradient} while the black dashed line is $\cos (\tau_i + \pi/5)$. }% The continuous line correspond to the best sinusoidal fit, $A\propto \cos(\phi+\pi/5)$.}
\label{fig:wavegradient_time_dependency}
\end{center}
\end{figure}

The amplitude of the Faraday waves generated by an impact is reported for different phases of impact $\tau_i = t_i \Omega / 2$ in figure~\ref{fig:wavegradient_time_dependency}. This amplitude is measured as the experimental envelope of $\partial_r \zeta(r,t)$ at a particular location $t$ and time $t$. It varies sinusoidally with $\tau_i$, so there are some impact phases for which the long-term excited wave is almost inexistent. Our model \eqref{eq:wave_gradient} predicts a dependence in $\cos(\tau_i-\pi/4)$ which is qualitatively similar. However, a dependence in $\cos(\tau_i + \pi/5)$ would correspond much better to the measurements. 

This phase shift between experiments and theory is likely due to the fact that equation~\eqref{eq:wave_gradient} represents an impulse response of surface waves, while in experiments the penetration of the bead into the bath is not instantaneous. 
According to Duhamel's principle, the response to the bead impact can be calculated by convolving the impulse response with the sophisticated pressure signal $\Pi_k(\tau)$ applied by the bead on the bath. This convolution is expected to approximately shift the effective impact time by half the time of penetration, which is in order of the contact time $5\sqrt{\rho R_d^3/\sigma}\sim 12~ms$ for a bead of radius $R_d = 0.5$~mm \citep{Molacek2013a}. On the other hand, the time delay corresponding to a phase shift from $-\pi/4$ to $\pi/5$ is 5.625 ms, which is very close to the 6 ms that such convolution would predict. %The amplitude after convolution should therefore vary as $\cos [\tau_i+\pi/5]$, which is in much better agreement with our experimental results. 

\section{Discussion and conclusion}\label{sec:Discussion}

In section~\ref{sec:theory}, we introduced a theory for the Faraday instability derived from Navier-Stokes equations and based on the only assumption that the surface waves are purely subharmonic. This assumption allows to truncate the Floquet series \eqref{eq:Ansatz_floquet}, and it is valid when $\Gamma \lesssim \Gamma_F$ (i.e. close to the Faraday threshold) for wavenumbers $k \simeq k_F$. We have then calculated the dispersion relation of Faraday waves, the Faraday threshold $\Gamma_F$ and the Faraday wavenumber, $k_F$. We have shown that for $\Gamma \lesssim \Gamma_F$, there is a range of $k$ in the vicinity of $k_F$, called the Faraday window, in which the frequency is locked to half the driving frequency (subharmonic response). Waves within this window are standing Faraday waves, while waves outside this window are dispersive travelling waves. The Faraday waves are long-lived, and their damping time is proportional to the so-called distance to threshold $\mathcal{M} = \Gamma_F / (\Gamma_F - \Gamma)$, as already observed in previous studies \citep{Eddi2011}. The present theory further predicts that the damping time is approximately inversely proportional to the fluid viscosity. 

The impulse response of the surface waves has been calculated in section~\ref{sec:ImpulseResponse}. Similarly to previous models (e.g. \citep{Eddi2011, Molacek2013b, Milewski2015}), its spatial dependence involves a Bessel function $J_0(k_F r)$ and its time evolution is a damped oscillation at frequency $\Omega/2$ (subharmonic response). Some of these previous models refer to a similar exponential damping in space. We show here that this damping is not strictly present in the impulse response. Instead, the wave amplitude is modulated by an exponential whose argument scales as $-r^2/(\tau - \tau_i)$. This modulation captures the diffusive propagation of the impulse excitation to the entire bath. A direct consequence is that two walkers do not instantly feel the impacts of each other. The time oscillation of the waves is phase-shifted by $\pi/4$ compared to the shaker signal. In other terms, the wave elevation $\zeta(r,t)$ is identically zero when the shaker is at its lowest position. This phase shift, confirmed experimentally, may significantly affect the vertical bouncing dynamics of the walker. Indeed, when the droplet impacts the bath, it feels an absolute acceleration that corresponds to the sum of contributions from the shaker motion and from the overlying phase-shifted Faraday waves. 

The wave field of a walker can be constructed by adding impulse responses corresponding to each impact of the droplet. This calculation has here been done in the case of a single walker moving at constant speed along a straight line. Under these assumptions, the vertical bouncing motion is perfectly periodic, and all impulse responses are given the same weight. However, the amplitude of the impulse response is modulated according to the phase of impact $\tau_i$ relative to the shaker, as predicted by this theory and confirmed experimentally. We expect the vertical dynamics of the walker to be perturbed as soon as additional Faraday waves are present. These Faraday waves may originate from the presence of other walkers, from the vicinity of a boundary, or from a previous passage of the walker at the same location (in the case of very high memory). In each of these cases, the vertical dynamics is not strictly periodic. Consequently, impact phase variations must be taken into account in the construction of the resulting wave field through an appropriately weighted summation of impulse responses. These refinements of the temporal description of Faraday waves are likely to be an additional source of chaos in the dynamics of walkers. 

The wave of a regular walker is shown to be at least exponentially damped in space. The characteristic damping length $l(\theta)$ is calculated (while it was experimentally estimated in previous works, e.g. in \citet{Eddi2011}). It increases with $\mathcal{M}$ and reaches a finite asymptotic value for $\mathcal{M} \rightarrow \infty$. It also decreases with increasing speed or viscosity. This damping length plays a crucial role for the walker dynamics, especially in the landmark double-slit experiment initially performed with walkers by \citet{Couder2006} and latter contested by \citet{Andersen2015}. In this experiment, a walking droplet passing to one slit has to possibly interfere with the waves that it sends to the other slit. Such interference is only possible if the waves emitted by the droplet before it crosses the slits are still significantly present when they are received by the droplet beyond the slits. This requires the damping time to be larger than the time required to cross the slits, and the damping length should then be significantly larger than the path length that they have to travel. If the slits are separated by a distance $d$, the damping length should then be larger than $\sqrt{2}d$. Experimental conditions of \citet{Couder2006, Andersen2015} are reported in table~\ref{tab:comparison}. As a matter of fact, none of them satisfied the two conditions for significant interferences: the damping time was always lower than the time to cross the slit, and the damping length was not much larger than $d$. In order to possibly observe interference patterns with single walkers, one would have to reproduce this experiment with a significantly higher memory (i.e. closer to Faraday threshold) and with a fluid of significantly lower viscosity.  
Similar conditions would have to be satisfied to perform experiments involving surreal trajectories.
\begin{table}
\begin{tabular}{lllllll}
\textsc{Experiment}&\hphantom{Latex} &$V_w$ (mm/s) & $\sqrt{2}d$ (mm)& $\mathcal{M}$& $\Me T_F V_w/ \sqrt{2}d$ & $l/\sqrt{2}d$ \\
&&&&&&\\
\citep{Couder2006}&&18.9&20.2&5--20& 0.12--0.5 & 0.8 \\
\citep{Andersen2015}&&15&14&21&0.5&1.17\\
%\citep{Pucci2017}&&12&28.3&500&5.3&2.18\\
\end{tabular}
\caption{Experimental conditions used in two previous double-slit experiments with walkers \citep{Couder2006, Andersen2015}. Walking speed $V_w$, path length of the waves $\sqrt{2}d$, distance to threshold $\mathcal{M}$, normalised damping time $\Me T_F V_w/\sqrt{2}d$ and normalised damping length $l / \sqrt{2}d$. }
\label{tab:comparison}
\end{table}

By contrast with walkers and Faraday waves, quantum mechanics is intrinsically non-dissipative. This feature can conceptually be approached with the walkers, by considering the limit of vanishing viscosity. In section~\ref{sec:lowvisc}, we have provided developments in series in the low viscosity limit. If viscosity vanished (i.e. here $\gamma_0 \rightarrow 0$), the Faraday threshold $\Gamma_F$ would vanish too, so the amplitude of the required driving vibration could be arbitrarily small. The memory $\Me$ would diverge and excitations from the bouncing droplets would be remembered by the wave forever. The diffusion coefficient $D$ and the damping length $l$ would diverge too, so each droplet impact would instantly modify the whole bath surface (with the compressibility of the liquid as a limit), i.e. each walker would instantly feel the presence of others, no matter how far they are. Therefore, in the limit of vanishing viscosity, walkers become intrinsically non-local objects composed of a local particle coupled to a non-local wave field. 

Walkers currently represent the only realisation at the macroscopic scale of a pilot-wave system similar to the one proposed by \citep{deBroglie1987} to interpret quantum mechanics \citep{Bush2015}. Many phenomena initially thought to be specific to the quantum realm have been observed with these walkers. The theory of Faraday waves developed in this paper brings a rigorous framework for the modelling of the wave field generated by a walker. It comprises additions to the previous Faraday wave models that are essential to the description of  walkers interfering with others, with boundaries, or with their past. It also provides insights on the locality of a walker, which has to be understood before experiments on the possible entanglement of two walkers could potentially be imagined. We hope that that this paper will help the building of a consistent wave theory to describe walkers similarly to Schr\"odinger's equation for quantum objects. It would then better reveal the limits of the analogy between walkers and the quantum world. 

\vspace{0.5cm}

%\section{Conclusion}

%In conclusion, we studied the Faraday instability for application to walkers. We found an elegant way for deriving the dispersion relation for surface waves under forcing, and thus approximate finely the Faraday threshold $\Gamma_F$ and the Faraday wavenumber, $k_F$. Those developments in series may be continued to get more accurate expressions for Faraday threshold. 
%
%For application to walkers, we moved away the study from the Faraday threshold for driving strengths just below the Faraday threshold. We showed that depending onto the distance to threshold $\mathcal{M}$, a Faraday window exists for wavenumbers close to the Faraday wavenumber $k_F$, in which the frequency is fixed to half the driving frequency. We found accordingly the damping time. We showed that the damping time varies proportionally to the distance to threshold $\mathcal{M}$ similarly to previous studies but with a dependency onto fluid viscosity. The wavefield generated by a walker is not exponentially damped in space but rather than driven by a exponential diffusive term. Faraday waves are also in advance of $\pi/4$ onto the driving strength.
%The theoretical results are supported with extended experimental measurements that confirms the theory. We also showed experimentally that the amplitude of Faraday waves depend onto impact phase.

\noindent\textbf{Acknowledgements}\vspace{0.2cm}

The authors thank Naresh Sampara and Brice Begasse for their help in building the experiment, as well as R\'emy Dubertrand and Ward Struyve for fruitful discussions. This work was financially supported by the `Actions de Recherches Concertées (ARC)' of the Belgium Wallonia Brussels Federation under contract No. 12-17/02 and by the FNRS grant number CHAR. RECH.-1.B423.18 (Tadrist L.).

%\appendix
%\section{}\label{appA}
\appendix
\section{Truncation of the Floquet ansatz}\label{app:Floquet}

The equation \eqref{eq:flf} of surface waves in the absence of external pressure $\Pi_k=0$ is recalled here:
\begin{equation}
f_k(s)\zeta_{k,s} + \frac{2\Gamma gk}{\Omega^2} \left(\zeta_{k, s+2i}+\zeta_{k, s-2i}\right) = 0 \,.
\label{eq:flfAPP}
\end{equation}
It can be solved with the Floquet ansatz \eqref{eq:Ansatz_floquetLaplace}: 
\begin{equation}
\zeta_{k,s} = \sum_{l=-\infty}^\infty \frac{\zeta_k^{(l)}}{s-(2l-1)i}
\label{eq:FloquetAnsatzAPP}
\end{equation}
which directly implies that
\begin{equation}
\quad\zeta_{k,s\pm 2i} = \sum_{l=-\infty}^\infty \frac{\zeta_k^{(l)}}{s\pm 2i-(2l-1)i} =  \sum_{l=-\infty}^\infty \frac{\zeta_k^{(l\mp 1)}}{s-(2l-1)i}\,.
\label{eq:FloquetAnsatzAPP2}
\end{equation}
Substitution of \eqref{eq:FloquetAnsatzAPP}~and~\eqref{eq:FloquetAnsatzAPP2} in \eqref{eq:flfAPP} yields
\begin{equation}
\sum_{l=-\infty}^\infty \frac{1}{s-(2l-1)i}\left(f_k(s) \zeta^{(l)}_k+ \frac{2\Gamma gk}{\Omega^2} \left(\zeta^{(l-1)}_k+\zeta^{(l+1)}_k\right)\right) = 0 \,.
\end{equation}
The Heaviside cover-up method is then applied to obtain a linear system of infinite dimension for the Floquet coefficients $\zeta_k^{(l)}$: 
\begin{equation}
\forall l \in \mathbb{Z},\quad f_k((2l-1)i) \zeta^{(l)}_k+ \frac{2\Gamma gk}{\Omega^2} \left(\zeta^{(l-1)}_k+\zeta^{(l+1)}_k\right)=0\,.
\end{equation}
of in matrix form
 \begin{equation}
 \underbrace{\left( \begin{array}{cccccccc}
\ddots & \ddots & \ddots& 0 & 0& 0&0&0 \\
0 & \alpha & f(-3i) &\alpha &0&0&0&0  \\
0 &0& \alpha & f(-i) &\alpha &0&0&0  \\
0 &0&0 &\alpha & f(i) &\alpha &0&0 \\
0 &0&0&0& \alpha & f(3i)&\alpha&0\\
0&0&0&0&0&\ddots&\ddots&\ddots   \end{array}  \right)}_K
 \left( \begin{array}{c}
 \vdots\\
 \zeta^{(-1)}\\
 \zeta^{(0)}\\
 \zeta^{(1)}\\
 \zeta^{(2)}\\
 \vdots\\
  \end{array}  \right) =\left( \begin{array}{c}
 \vdots\\
 0\\
 0\\
 0\\
 0\\
 \vdots\\
  \end{array}  \right)
 \end{equation}
where we noted $\alpha = 2\Gamma g k /\Omega^2$ for readability. The matrix is tridiagonal, which indicates that each Floquet mode is coupled to its neighbour modes. The condition of existence of surface waves is bound to the obtention of a non-trivial solution to this system. This can only be achieved when the determinant $\det K = 0$, which yields the threshold $\Gamma_k$ associated to each wavenumber $k$ for the non-truncated Floquet ansatz. 

In the main text, we truncated the series to the two first modes, which led to equations~\eqref{eq:floquet1}~and~\eqref{eq:floquet2}. The approximation of this truncation may be evaluated by considering the presence of the two next modes. With a truncation enlarged to four modes, the linear system becomes
\begin{equation}
 \left( \begin{array}{cccc}
 f_k(-3i) &\alpha &0&0  \\
\alpha & f_k(-i) &\alpha &0 \\
0 &\alpha & f_k(i) &\alpha  \\
0&0& \alpha & f_k(3i)  \end{array}  \right)
 \left( \begin{array}{c}
 \zeta^{(-1)}\\
 \zeta^{(0)}\\
 \zeta^{(1)}\\
 \zeta^{(2)}\\
  \end{array}  \right) =\left( \begin{array}{c}
 0\\
 0\\
 0\\
 0\\
  \end{array}  \right)
\end{equation}
A non-trivial solution is found when the determinant is zero, i.e. 
\begin{equation}
\frac{\alpha^4}{|f_k(3i)|^2} - \alpha^2\left(1+\frac{f_k(-i)}{f_k(3i)}+\frac{f_k(i)}{f_k(-3i)}\right) +|f_k(i)|^2=0
\end{equation}
The curve $\Gamma_k$ is given by the smallest positive solution to this equation, namely,
\begin{equation}
\frac{\alpha}{|f_k(3i)|} = \frac{1}{\sqrt{2}}\sqrt{\left(1+\frac{f_k(-i)}{f_k(3i)}+\frac{f_k(i)}{f_k(-3i)}\right) - \sqrt{\left(1+\frac{f_k(-i)}{f_k(3i)}+\frac{f_k(i)}{f_k(-3i)}\right)^2-4\frac{|f_k(i)|^2}{|f_k(3i)|^2}}}
\label{eq:exact_sol_floquet}
\end{equation}
Around $k_F$, in the low viscosity limit (cf. appendix~\ref{app:low_viscosity}), one can show that for $k \sim k_F$, $|f_k(i)|^2 \sim 4\gamma_k^2$ and $|f_k(3i)|_F^2\sim 64$. Consequently, $|f(i)|_F^2/|f(3i)^2|_F \sim \gamma_F^2/16 \ll 1$. Equation~\eqref{eq:exact_sol_floquet} can then be expanded in Taylor series, which yields
\begin{equation}
\Gamma_k = \frac{\Omega^2|f_k(i)| }{ 2 g k}\left( 1-\frac{|f_k(i)|^2}{2|f_k(3i)|^2}\frac{1}{(1+f_k(-i)/f_k(3i)+f_k(i)/f_k(-3i))^2}\right) \, .
\label{eq:exact_sol_floquet2}
\end{equation}
The correction to $\Gamma_k$ corresponding to the addition of four terms instead of two in the Floquet ansatz is of the order of $|f(i)|^2/2|f(3i)|^2\sim \gamma_F^2/32 < 1\%$ in the limit of small viscosity and for $k \sim k_F$. 

\begin{figure}
\begin{center}
\begin{psfrags}
\psfrag{f}[c][c]{Faraday}\psfrag{w}[c][c]{window}
\psfrag{fw}[c][c]{}\psfrag{dk}[c][c]{$\delta_k$}
\psfrag{dk1}[c][c]{$\delta_k^+$}\psfrag{dk2}[c][c]{$\delta_k^-$}
\psfrag{gk}[c][c]{$\Gamma_k$}\psfrag{k}[c][c]{$k$ (m$^{-1}$)}
\psfrag{kF}[c][c]{$k_F$}\psfrag{gF}[c][c]{$\Gamma_F$}
\psfrag{Im}[l][l]{$\mathcal{I}_m(i+\delta_k)$}
\psfrag{Im1}[l][l][0.8]{$\mathcal{I}_m(i+\delta_k^+)$}
\psfrag{Im2}[l][l][0.8]{$\mathcal{I}_m(i+\delta_k^-)$}
\psfrag{4}[c][c][0.8]{4}\psfrag{6}[c][c][0.8]{6}\psfrag{8}[c][c][0.8]{8}\psfrag{5}[c][c][0.8]{5}\psfrag{7}[c][c][0.8]{7}
\psfrag{1000}[c][c][0.8]{1000}\psfrag{1200}[c][c][0.8]{1200}\psfrag{1400}[c][c][0.8]{1400}\psfrag{1600}[c][c][0.8]{1600}\psfrag{1800}[c][c][0.8]{1800}
\psfrag{-0.4}[c][c][0.8]{-0.4}\psfrag{-0.3}[c][c][0.8]{-0.3}\psfrag{-0.2}[c][c][0.8]{-0.2}\psfrag{-0.1}[c][c][0.8]{-0.1}\psfrag{0}[c][c][0.8]{0}\psfrag{0.6}[cb][c][0.8]{0.6}\psfrag{0.8}[c][c][0.8]{0.8}\psfrag{1}[c][c][0.8]{1}\psfrag{1.2}[c][c][0.8]{1.2}\psfrag{1.4}[c][c][0.8]{1.4}
\includegraphics[width = 0.44\textwidth]{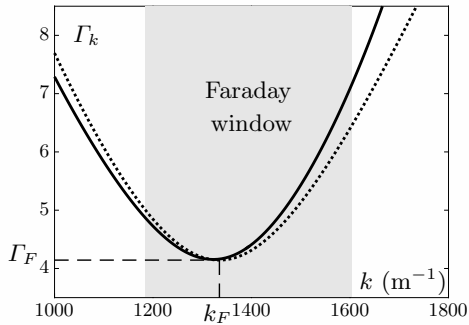}
\end{psfrags}
\end{center}
\caption{Faraday threshold $\Gamma_k$ of each wavenumber $k$. The dashed line represents a truncation to the two first modes. The solid line corresponds to the truncation to either four or twenty modes - both being indistinguishable from each other.}
\label{fig:floquet_truncation}
\end{figure}

Figure~\ref{fig:floquet_truncation} represents the curve $\Gamma_k$, calculated with a truncation to 2 modes, to 4 modes and to 20 modes respectively. The first two modes already provide a very good estimation of the curve $\Gamma_k$, especially for $k \simeq k_F$. The curves $\Gamma_k$ for four and twenty modes are almost indistinguishable. For walker parameters (cf. table~\ref{tab:notations}), $\Gamma_F=4.155$ with a truncation to 20 modes whereas $\Gamma_F= 4.144$ with a truncation to only 2 modes. Similarly, the Faraday wavelength is $\lambda_F = 4.749$~mm with 20 modes, instead of $\lambda_F = 4.706$ with two modes. In both cases, the relative error from the truncation to two modes is less than 1\%. 

\section{Low-viscosity asymptotic behaviour}\label{app:low_viscosity}

\subsection{Preliminary calculations}

The function $f_k$ and its derivatives are first recalled:
 \begin{eqnarray}
  f_k(s) &=& (s+\gamma_k)^2+\omega_k^2 - \gamma_k^{3/2}\sqrt{\gamma_k+2s}\,, \\ 
  \dot f_k(s) &=& 2(s+\gamma_k)-\frac{\gamma_k^{3/2}}{\sqrt{\gamma_k+2s}}\,, \\
  \ddot f_k(s) &=& 2 +\left(\frac{\gamma_k}{\gamma_k+2s}\right)^{3/2}\,.
 \end{eqnarray}
They are all self-conjugated, so e.g. $f_k(s^*) = f_k(s)^*$. 

In order to evaluate them in $s = \pm i$, we first calculate 
\begin{equation}
\sqrt{\gamma_k\pm2i} =\frac{1}{\sqrt{2}}\sqrt{\sqrt{\gamma_k^2+4}+\gamma_k}\pm\frac{i}{\sqrt{2}}\sqrt{\sqrt{\gamma_k^2+4}-\gamma_k}
\end{equation} 
Only the positive real part is considered, since the vertical fluid velocity must remain finite when $z\longrightarrow -\infty$ in equation~\eqref{eq:vertical_velocity_solution}.
   
Therefore, without any approximation, 
\begin{eqnarray}
\mathbb{R}(f_k(i)) &=& \gamma_k^2+\omega_k^2-1-\frac{\gamma_k^{3/2}}{\sqrt{2}}\sqrt{\sqrt{\gamma_k^2+4}+\gamma_k}\\
\mathbb{I}(f_k(i)) &=& 2\gamma_k -\frac{\gamma_k^{3/2}}{\sqrt{2}}\sqrt{\sqrt{\gamma_k^2+4}-\gamma_k} \\
\mathbb{R}(\dot f_k(i)) &=& 2\gamma_k-\frac{\gamma_k^{3/2}}{\sqrt{2}\sqrt{\gamma_k^2+4}}\sqrt{\sqrt{\gamma_k^2+4}+\gamma_k}\\
\mathbb{I}(\dot f_k(i)) &=& 2 +\frac{\gamma_k^{3/2}}{\sqrt{2}\sqrt{\gamma_k^2+4}}\sqrt{\sqrt{\gamma_k^2+4}-\gamma_k}\\
\mathbb{R}(\ddot f_k(i)) &=& 2+\frac{\gamma_k^{3/2}}{\sqrt{2}(\gamma_k^2+4)^{3/2}}\left(\gamma_k\sqrt{\sqrt{\gamma_k^2+4}+\gamma_k}-2\sqrt{\sqrt{\gamma_k^2+4}-\gamma_k}\right)\\
\mathbb{I}(\ddot f_k(i)) &=& -\frac{\gamma_k^{3/2}}{\sqrt{2}(\gamma_k^2+4)^{3/2}}\left(\gamma_k\sqrt{\sqrt{\gamma_k^2+4}-\gamma_k}+2\sqrt{\sqrt{\gamma_k^2+4}+\gamma_k}\right)
\end{eqnarray} 
 
When the viscosity is sufficiently small, $\gamma_k \ll 1$, and both $f_k(i)$ and its derivatives can be approximated with a Taylor series in $\gamma_k$: 
% with $\sqrt{\sqrt{\gamma_k^2+4}+\gamma_k} \sim \sqrt{2}(1+\gamma_k/4)$ and $\sqrt{\sqrt{\gamma_k^2+4}+\gamma_k}\sim \sqrt{2}(1-\gamma_k/4)$,
 \begin{eqnarray}
\mathbb{R}(f_k(i)) &=& \omega_k^2-1-\gamma_k^{3/2}+\gamma_k^2-\frac{1}{4}\gamma_k^{5/2}+\mathcal{O}(\gamma_k^{7/2})\\
\mathbb{I}(f_k(i)) &=& 2\gamma_k -\gamma_k^{3/2}+\frac{1}{4}\gamma_k^{5/2}+\mathcal{O}(\gamma_k^{7/2})\\
\mathbb{R}(\dot f_k(i)) &=& 2\gamma_k-\frac{\gamma_k^{3/2}}{2}-\frac{\gamma_k^{5/2}}{8}+\mathcal{O}(\gamma_k^{7/2})\\
\mathbb{I}(\dot f_k(i)) &=&  2+\frac{\gamma_k^{3/2}}{2}-\frac{\gamma_k^{5/2}}{8}+\mathcal{O}(\gamma_k^{7/2})\\
\mathbb{R}(\ddot f_k(i)) &=& 2-\frac{\gamma_k^{3/2}}{4}+\frac{3}{16}\gamma_k^{5/2}+\mathcal{O}(\gamma_k^{7/2})\\
\mathbb{I}(\ddot f_k(i)) &=& -\frac{\gamma_k^{3/2}}{4}-\frac{3}{16}\gamma_k^{5/2}+\mathcal{O}(\gamma_k^{7/2})
\end{eqnarray} 
 
The following combinations of $f_k(i)$ and its derivatives are regularly called in section~\ref{sec:theory}:
 \begin{equation}
|f_k(i)|^2 = (\omega_k^2-1)^2-2(\omega_k^2-1)\gamma_k^{3/2}+2(\omega_k^2+1)\gamma_k^2-\left(4+\frac{\omega_k^2-1}{2}\right)\gamma_k^{5/2}+2\gamma_k^3+\mathcal{O}(\gamma_k^{7/2})\, ,
\label{eq:estimatefi2}
\end{equation}
\begin{equation}
f_k(i)\dot f_k(-i)+f_k(-i)\dot f_k(i) = 4\gamma_k(\omega_k^2+1) - (\omega_k^2+3)\gamma_k^{3/2} - \left(\frac{\omega_k^2+3}{4}\right)\gamma_k^{5/2} + \mathcal{O}(\gamma_k^3)\, ,
\label{eq:estimate_fifdot-i}
\end{equation}
\begin{equation}
\frac{f_k(i)\ddot f_k(-i) + f_k(-i)\ddot f_k(i)}{2} + \dot f_k(i)\dot f_k(-i) = 2(\omega_k^2+1) - \frac{\omega_k^2-1}{4} {\gamma_k}^{3/2}+ 6\gamma_k^2 +\frac{3\omega_k^2-59}{16}\gamma_k^{5/2}+ \mathcal{O}(\gamma_k^{7/2})
\end{equation}

Finally, the two first $k$-derivative of $|f_k(i)|^2$, useful for the calculation of the Faraday threshold and the diffusion coefficient, are expanded as:
\begin{eqnarray}
\nonumber \frac{k}{2}\frac{\partial |f_k(i)|^2}{\partial k} &=& 2k(\omega_k^2-1)\omega_k c_k -\left(2k\omega_k c_k+3(\omega_k^2-1)\right)\gamma_k^{3/2} + \left(2k\omega_k c_k+4(\omega_k^2+1)\right)\gamma_k^2\\
& &-\left[\frac{k\omega_k c_k}{2}+\frac{5}{2}\left(4+\frac{\omega_k^2-1}{2}\right)\right]\gamma_k^{5/2}+6\gamma_k^3 +\mathcal{O}(\gamma_k^{7/2}) \label{eq:estimatedfi2} \\
\frac{\partial^2 |f_k(i)|^2}{\partial k^2} &=& 8 \omega_k^2 c_k^2 + 4 (\omega_k^2 - 1) (c_k^2 + \omega_k \dot{c}_k) \nonumber \\
& & - \left( 4 (c_k^2 + \omega_k \dot{c}_k) + \frac{12}{k^2} (\omega_k^2-1) + \frac{24 \omega_k c_k}{k}\right) \gamma_k^{3/2} + \mathcal{O}(\gamma_k^{2})
\end{eqnarray}
where $c_k = \partial\omega_k/\partial k$ is the dimensionless group velocity, and $\dot{c_k} = \partial c_k / \partial k$. 

The low viscosity approximation of the damping of Faraday waves reads 
\begin{equation}
\delta_k^+ = -\frac{ \left((\omega_k^2-1)^2-2(\omega_k^2-1)\gamma_k^{3/2}+2(\omega_k^2+1)\gamma_k^2-\left(4+\frac{\omega_k^2-1}{2}\right)\gamma_k^{5/2}-(2 \Gamma g k/\Omega^2)^2\right)}{4(\omega_k^2+1)\gamma_k -(\omega_k^2+3) \gamma_k^{3/2}-(\omega_k^2+3)\gamma_k^{5/2}/4}+\mathcal{O}(\gamma_k^{3})
 \label{eq:damping_k_approx}
\end{equation}

\subsection{Determination of $k_F$, $\omega_F$ and $c_F$}

The most unstable wave number $k_F$ is defined as the solution of 
\begin{equation}
\frac{k}{2}\frac{\partial |f_k(i)|^2}{\partial k} = |f_k(i)|^2\, .
\end{equation}
According to the approximations \eqref{eq:estimatefi2} and \eqref{eq:estimatedfi2}, it satisfies
\begin{eqnarray}
\nonumber& &(\omega^2_F-1)(\omega^2_F-1-2{k_F}\omega_F c_F)+ (\omega^2_F-1+2{k_F}\omega_F c_F)\gamma_F^{3/2}\\ \nonumber &-& \left(2(\omega^2_F+1)+2{k_F}\omega_F c_F\right)\gamma_F^2+\left(6+\frac{3}{4}(\omega^2_F-1)+\frac{{k_F}\omega_F c_F}{2}\right)\gamma_F^{5/2}\\&-&4\gamma_F^3+\mathcal{O}(\gamma_F^{7/2})=0
\label{eq:ExpansionkF}
\end{eqnarray}
where $\omega_F = \omega_{k_F}$ and $c_F = c_{k_F}$. 

When the viscosity $\nu$ is identically zero, $\gamma_F = 0$ and equation~\eqref{eq:ExpansionkF} is simplified in
\begin{equation}
(\omega_F^2 - 1) (\omega_F^2 - 1 - 2 k_F \omega_F c_F) = 0
\end{equation}
The second factor is strictly negative for $k_F > 0$, so $k_F$ must satisfy $\omega_F^2 = 1$. We define $$k_0 = \lim_{\nu \rightarrow 0} k_F$$ for which $\omega_0 = \omega_{k_0} = 1$, $c_0 = c_{k_0}$ and $\gamma_0 = \gamma_{k_0}$. 

The damping factor $\gamma_0 = 4 \nu k_0^2 / \Omega$ is a better basis for an asymptotic expansion at low viscosity, since it does not directly depend on the Faraday wavenumber $k_F$, which is still unknown. Equation~\eqref{eq:ExpansionkF} can be rewritten in terms of $\gamma_0$ since $\gamma_F = k_F^2\gamma_0/k_0^2$: 

\begin{eqnarray}
\nonumber& &(\omega^2_F-1)(\omega^2_F-1-2{k_F}\omega_F c_F)+ (\omega^2_F-1+2{k_F}\omega_F c_F)\left(\frac{k_F}{k_0}\right)^3\gamma_0^{3/2}\\ \nonumber &-& \left(2(\omega^2_F+1)+2{k_F}\omega_F c_F\right)\left(\frac{k_F}{k_0}\right)^4\gamma_0^2+\left(6+\frac{3}{4}(\omega^2_F-1)+\frac{{k_F}\omega_F c_F}{2}\right)\left(\frac{k_F}{k_0}\right)^5\gamma_0^{5/2}\\&-&4\left(\frac{k_F}{k_0}\right)^6\gamma_0^3+\mathcal{O}(\gamma_0^{7/2})=0\,.
\label{eq:explication_iterative_k_F}
\end{eqnarray}

We formulate the ansatz $$k_F= k_0 +\epsilon \gamma_0^{\alpha} + o(\gamma_0^{\alpha}),$$ where $\alpha >0$, from which we directly infer that $$\omega_F^2 = 1 + 2 c_0 \epsilon \gamma_0^{\alpha} + o(\gamma_0^{\alpha})$$ and $$2 \omega_F c_F = 2 c_0 + \frac{24 \sigma}{\rho \Omega^2} k_0 \epsilon \gamma_0^{\alpha} + o(\gamma_0^{\alpha}).$$ Inserting this ansatz in equation~\eqref{eq:explication_iterative_k_F} yields
\begin{equation}
2 c_0 k_0 \left( \gamma_0^{3/2} - 2 c_0 \epsilon \gamma_0^{\alpha} \right) + o(\gamma^{\alpha}) + o(\gamma^{3/2}) = 0 
\end{equation}
from which we deduce $\alpha = 3/2$ and $\epsilon = 1/(2 c_0)$, so $k_F = k_0 + \frac{1}{2c_0} \gamma_0^{3/2} + o(\gamma_0^{3/2})$. We can iterate this procedure with a new ansatz $k_F= k_0 + \gamma_0^{3/2}/(2c_0) + \epsilon \gamma_0^{\alpha} + o(\gamma_0^{\alpha})$ in order to find higher-order corrections. It yields:
\begin{eqnarray}
k_F&=& k_0 + \frac{1}{2c_0}\gamma_0^{3/2}-\frac{2+k_0c_0}{2k_0 c_0^2}\gamma_0^{2}+\frac{12+k_0 c_0}{8 k_0 c_0^2}\gamma_0^{5/2}+\left(\frac{1}{4k_0 c_0^2}-\frac{d_0}{8c_0^3}\right)\gamma_0^3+\mathcal{O}(\gamma_0^{7/2}) \nonumber \\
\omega_F^2 &=& 1 + \gamma_0^{3/2} -\frac{2+k_0 c_0}{k_0 c_0}\gamma_0^2+ \frac{12+k_0 c_0}{4 k_0 c_0} \gamma_0^{5/2} +\mathcal{O}\left(\gamma_0^3\right) \nonumber \\
\gamma_F &=& \gamma_0+\frac{\gamma_0^{5/2}}{k_0 c_0}+\mathcal{O}\left(\gamma_0^3\right) \, .
\end{eqnarray}
where  $d_0=(\partial^2\omega_k^2/\partial k^2) (k_0)$.

From there, we evaluate:
 \begin{eqnarray}
 \mathbb{R}(f_F(i)) &=& \frac{-2}{k_0 c_0}\gamma_0^2+\frac{3}{k_0 c_0}\gamma_0^{5/2}+\mathcal{O}\left(\gamma_0^3\right) \nonumber \\
 \mathbb{I}(f_F(i)) &=& 2\gamma_0-\gamma_0^{3/2}+\frac{8+k_0 c_0}{4 k_0 c_0}\gamma_0^{5/2}+\mathcal{O}\left(\gamma_0^3\right) \nonumber \\
f_F(i)\dot f_F(-i)+f_F(-i)\dot f_F(i) &=& 8\gamma_0-4\gamma_0^{3/2}+\mathcal{O}\left(\gamma_0^2\right) \nonumber \\
\frac{f_F(i)\ddot f_F(-i) + f_F(-i)\ddot f_F(i)}{2} + \dot f_F(i)\dot f_F(-i) &=& 2\gamma_0^{3/2}+4\left(1-\frac{1}{k_0 c_0}\right)\gamma_0^2 \nonumber \\
&+&\frac{12+9k_0 c_0 }{2k_0 c_0}\gamma_0^{5/2}+\mathcal{O}\left(\gamma_0^3\right) \nonumber \\
\left.\frac{\partial^2|f_k(i)|^2}{\partial k^2}\right|_F &=& 8 c_0^2 + \mathcal{O}\left(\gamma_0^{3/2}\right) 
 \end{eqnarray}

%From equations \eqref{eq:delta+dev} and \eqref{eq:D}, it comes directly that
%\begin{equation}
%D=\frac{\Gamma_F(2 g k_F/\Omega^2)^2}{2\left(f_F(i)\dot{f}_F(-i)+f_F(-i)\dot{f}_F(i)\right)}\left(\frac{\partial^2\Gamma_{k}}{\partial k^2}\right)_F
%\end{equation}
%We have now to evaluate $\partial^2\Gamma_{k}/\partial k^2$ in $k_F$. To this aim, we remark that
%\begin{equation}
%\left(\frac{\partial^2\Gamma_{k}}{\partial k^2}\right)_F = \frac{1}{2k_F^2\Gamma_F}\left.\frac{\partial^2\left(k\Gamma_k\right)^2}{\partial k^2}\right|_{F} - \frac{\Gamma_F}{k_F^2}
%\end{equation} 
%since $(\partial \Gamma_k/\partial k)_F=0$ by definition of $k_F$. With the relation $k^2\Gamma_k^2 = \Omega^4|f_k(i)|^2/(2g)^2$, we obtain the exact expression of $D$,

\section{Inverse Hankel transform}\label{app:hankel}

In this appendix, we evaluate the integral \eqref{eq:zeta1}, namely
\begin{equation}
\zeta_1(r,\tau) = \int_0^\infty B_k^+ \cos (\tau + \theta_k^+) e^{\delta_k^+ (\tau - \tau_i)} J_0(kr) k dk \, ,
\end{equation}
with the saddle-node method. First, $\delta_k^+$ is replaced by its Taylor series around $k_F$: 
\begin{equation}
\delta_k^+ \simeq \delta_F - D (k-k_F)^2 \, .
\end{equation}
Substitution in the integral yields
\begin{equation}
\zeta_1(r,\tau) = e^{\delta_F (\tau - \tau_i)} \int_0^\infty B_k^+ \cos (\tau + \theta_k^+) e^{-D (\tau - \tau_i)(k-k_F)^2} J_0(kr) k dk \, ,
\end{equation}

In the limit $k_F r \ll 1$, the Bessel function does not oscillate significantly. Then for $\tau \gg \tau_i$, 
\begin{eqnarray}
\zeta_1(r,\tau) &\simeq& e^{\delta_F (\tau - \tau_i)} B_F^+  \cos (\tau + \theta_F^+) J_0(k_F r) k_F \int_0^{\infty} e^{-D(\tau-\tau_i)(k-k_F)^2} \mathrm{d}k \nonumber \\
&=& B_F^+ k_F  e^{\delta_F (\tau - \tau_i)}  \cos (\tau + \theta_F^+) \sqrt{\frac{\pi}{D(\tau-\tau_i)}}
J_0(k_F r) \label{eq:app_henkel_small_r}
\end{eqnarray}

For $k_F r \gg 1$, the approximation of $J_0$ by $\mathbb{R} \lbrace e^{ikr+i\pi/4}/\sqrt{\pi kr/2}\rbrace$ yields 
\begin{eqnarray}
\zeta_1(r,\tau) &\simeq& e^{\delta_F (\tau - \tau_i)} \int_0^\infty B_k^+ \cos (\tau + \theta_k^+) e^{-D (\tau - \tau_i)(k-k_F)^2} 
e^{ikr+i\pi/4} \sqrt{\frac{2}{\pi kr}} k \mathrm{d}k \nonumber \\
&\simeq& e^{\delta_F (\tau - \tau_i)} B_F^+ \cos (\tau + \theta_F^+) J_0(k_F r) k_F  \int_0^\infty e^{-D (\tau - \tau_i)(k-k_F)^2} e^{i(k-k_F)r} \mathrm{d}k
\end{eqnarray}
The remaining integral is calculated as
\begin{eqnarray}
\int_0^\infty e^{-D (\tau - \tau_i)(k-k_F)^2} e^{i(k-k_F)r} \mathrm{d}k &=& \exp\left[- \frac{r^2}{4 D (\tau - \tau_i)}\right]  \int_0^{\infty}  e^{-D(\tau - \tau_i) \left[ k - k_F - \frac{i r}{2 D (\tau - \tau_i)}\right]^2} \mathrm{d}k \nonumber \\
&=&  \exp\left[- \frac{r^2}{4 D (\tau - \tau_i)}\right]  \sqrt{\frac{\pi}{D(\tau-\tau_i)}}
\end{eqnarray}
Therefore, 
\begin{equation}
\zeta_1(r,\tau) \simeq  B_F^+ k_F e^{\delta_F (\tau - \tau_i)} \cos (\tau + \theta_F^+) \sqrt{\frac{\pi}{D(\tau-\tau_i)}}
 J_0(k_F r) \exp\left[- \frac{r^2}{4 D (\tau - \tau_i)}\right] 
 \label{eq:app_hankel_large_r}
 \end{equation}
Coincidentally, this asymptotic behaviour at $k_F r \gg 1$ is also valid for $k_F r \ll 1$. Indeed, equations \eqref{eq:app_henkel_small_r} and \eqref{eq:app_hankel_large_r} are identical as soon as $\tau-\tau_i\gg r^2 / (4 D)$. This condition is necessarily satisfied as soon as $\tau - \tau_i \gg 1 / (4 D k_F^2)$ for $k_F r \ll 1$. For walkers, with numerical values in table \ref{tab:notations}, $1/4Dk_F^2 \sim 0.08$ so equation~\eqref{eq:app_hankel_large_r} is equivalently valid at $k_F r \gg 1$ and $k_F r \ll 1$.

\section{Surface waves emitted by a walker}\label{app:walker}

The Faraday wave profile resulting from the previous bounces of the walker at the positions $\mathbf{r}_n = - (\tau - \tau_n)\mathbf{v}$ and the times $\tau_n = \tau_i + 2\pi n$ ($n\in\mathbb{Z}$) reads
\begin{eqnarray}
\zeta_w(\mathbf{r},\tau) &=& \sum_{n=-\infty}^{[(\tau-\tau_i)/2\pi]} \zeta_1 (|\mathbf{r}-\mathbf{r}_n|,\tau-\tau_n)\nonumber\\
&=& \sum_{n=-\infty}^{[(\tau-\tau_i)/2\pi]} B_F^+ k_F\sqrt{\frac{\pi}{D(\tau - \tau_n)}} J_0(k_F |\mathbf{r}-\mathbf{r}_n|) \cos\left(\tau+\theta_F^+ \right)\nonumber\\
&\times & \exp\left(-\frac{\tau-\tau_n}{2\pi \Me}-\frac{|\mathbf{r}-\mathbf{r}_n|^2}{4D(\tau-\tau_n)}\right) \,, \label{eq:swp}
\end{eqnarray}
where $[(\tau - \tau_i)/2\pi]$ denotes the largest integer smaller than $(\tau - \tau_i)/2\pi$, and where the origin $\mathbf{r} = 0$ is placed at the position of the walker at time $\tau$. We replace the sum in the previous equation by an integration over the delay time $u = \tau - \tau_n$ going from $0$ to $\infty$. This is justified for $\Me r^2 \gg 2\pi D$, since the sum can then be seen as an approximate Riemann integral where the interpolated function does not significantly vary over one bin.

The Graf's theorem for $0^\mathrm{th}$-order Bessel functions reads
\begin{equation}
J_0(k_F |\mathbf{r}-\mathbf{r}_n|) = \sum_{n=-\infty}^\infty J_n(k_F r) J_n(k_F r_n) e^{-i n \theta}
\end{equation}
with
\begin{equation}
  \theta = \arccos\left( \frac{\mathbf{v}\cdot\mathbf{r}}{v r} \right) \,.
\end{equation}
Since Bessel functions are defined as
\begin{equation}
J_n(k_F r) = \frac{1}{2\pi}\int_{-\pi}^{\pi} \mathrm{d}\varphi\, e^{i k_F r \sin\varphi}\, e^{-in\varphi}
\end{equation}
and since $\sum\left[ e^{i(\theta-\varphi-\varphi')}\right]^n = \delta(\theta-\varphi-\varphi')$, it results that 
\begin{equation}
  J_0(k_F |\mathbf{r}-\mathbf{r}_n|) = \frac{1}{2\pi} \int_{-\pi}^\pi 
  \exp[i k_F r \sin(\theta - \varphi)] \exp(- i k_F r_n \sin\varphi)\,d\varphi
  \label{eq:graf}
\end{equation}

Substitution in equation~\eqref{eq:swp} yields
\begin{eqnarray}
\zeta_w(\mathbf{r},\tau) &=& \frac{B_F^+ k_F}{2 \sqrt{\pi D}} \cos\left(\tau+\theta_F^+\right) \int_{-\pi}^\pi d\varphi \exp[i k_F r \sin(\theta - \varphi)] \\
&\times & \int_0^\infty\frac{1}{\sqrt{u}} \exp\left(-\frac{u}{2\pi M\mathrm{e}}-\frac{r^2}{4Du} -\frac{v r\cos \theta}{2 D} - \frac{v^2 u}{4 D} + i k_F u v \sin\varphi\right)\mathrm{d}u \,. \nonumber
\end{eqnarray}
Using the identity
\begin{equation}
\int_0^\infty \frac{1}{\sqrt{u}} \exp\left(-\alpha u -\beta /u\right)\mathrm{d} u = \sqrt{\frac{\pi}{\alpha}}\exp\left(-2\sqrt{\alpha\beta}\right)\\
\end{equation}
for $\mathbb{R}(\alpha)>0$ and $\beta>0$, we obtain
\begin{eqnarray}
\zeta_w(\mathbf{r},\tau) &=& \frac{B_F^+ k_F}{2\sqrt{D}} \cos\left(\tau+\theta_F^+ \right) \exp\left(- \frac{v r \cos\theta}{2D} \right) \\
&\times& \int_{-\pi}^\pi d\varphi \frac{\exp[i k_F r \sin(\theta - \varphi)]}
{\sqrt{a- i k_F v \sin\varphi}} 
\exp{\left(-\frac{r}{\sqrt{D}}\sqrt{a - i k_F v \sin\varphi}\right)} \nonumber
\end{eqnarray}
where 
\begin{equation}
  a = \frac{1}{2\pi\Me} + \frac{v^2}{4D} \,.
\end{equation}

For large $a$ such that $ k_F v / a \ll 1$ (for walkers, with numerical values in table \ref{tab:notations}, this correspond to a distance to threshold $\mathcal{M} < 4.8$), we can justify the approximation
\begin{equation}
\sqrt{a - i k_F v \sin\varphi} \simeq \sqrt{a} - i\frac{k_F v }{2\sqrt{a}} \sin\varphi \,.
\end{equation}
A reverse application of Graf's theorem (equation~\eqref{eq:graf}) yields
\begin{eqnarray}
\zeta_w(\mathbf{r},\tau) &=& \frac{B_F^+ k_F}{2\sqrt{D a}} \cos\left(\tau+\theta_F^+ \right) J_0\left(k_F\left| \mathbf{r} + \frac{r}{2\sqrt{Da}} \mathbf{v} \right|\right) \nonumber \\
&\times &\exp\left[- \left( \sqrt{\frac{a}{D}} + \frac{v \cos\theta}{2D} \right) r \right] \,,
\end{eqnarray}
i.e., we obtain a Doppler-shifted Bessel profile which is exponentially attenuated in space. The damping length scale, defined as $l(\theta)$, depends on the orientation $\theta$ with respect to the walking direction ${\bf v}/v$: 
\begin{equation}
  l(\theta) = \frac{2D}{v\cos\theta + \sqrt{v^2+2D/\pi\Me}} \,. \label{eq:l}
\end{equation}
In the special case of a stationary bouncer with $v=0$, we obtain the spatial damping length $l = \sqrt{2\pi\Me D}$ which diverges in the limit $\Me\to\infty$.

An accurate analytical evaluation of equation~\eqref{eq:walker} in the more general case $ k_F v/ a \gtrsim 1$ is more involved and beyond the scope of this work. However, we can straightforwardly provide an upper limit for the wave amplitude which holds in any case, namely through the inequality
\begin{equation}
\left| \frac{\zeta_w(\mathbf{r},\tau)}{\cos(\tau+\theta_F^+)}\right| \leq \frac{\pi B_F^+ k_F}{\sqrt{D a}} \exp\left[-\frac{r}{l(\theta)}\right] \,,
\end{equation}
obtained by substituting $|J_0(k_F |\mathbf{r}-\mathbf{r}_n|)| \leq 1$ within equation~\eqref{eq:swp}.

\section{Experimental measurement of the impact phase}\label{app:experimental_set_up}

\begin{figure}
\begin{center}
\begin{psfrags}
\psfrag{S}[l][l]{\hspace{-0.5cm}Shaker} \psfrag{p}[l][l]{\hspace{-0.5cm}acceleration}\psfrag{i}[c][c]{impact}\psfrag{ti}[c][c]{$t_i$}\psfrag{time}[c][c]{time}\psfrag{F}[c][c]{$s$}\psfrag{X}[c][c]{$t_f$}\psfrag{o}[c][c]{\textcolor{red}{0}}\psfrag{rt}[c][c]{recording period}\psfrag{o}[c][c]{$0$}\psfrag{n}[c][c]{\textcolor{red}{$n\,T_F$}}\psfrag{k}[c][c]{$\Delta t$}
\includegraphics[width=0.6\textwidth]{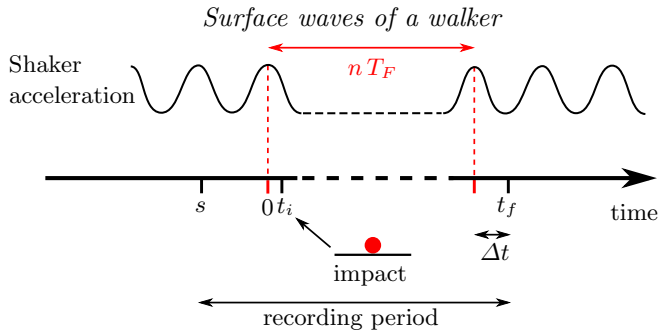}
\end{psfrags}
\caption{Measurement of the impact time $t_i$, from the end time of the recording $t_f$ and the synchronisation delay $\Delta t$. Definitions of $t_i$, $t_f$ and $\Delta t$ are provided in the main text.}
\label{fig:synchro}
\end{center}
\end{figure}

Measuring the impact phase requires a good synchronisation of the high-speed camera with the vibration $g(t)$ of the shaker. For this purpose, the trigger of the camera that marks the end of the recording is also sent to the DACQ. Both signals are simultaneously processed with a \textit{labview} interface that outputs the time delay $\Delta t \in [0, T_F/2]$ from the last recorded maximum of $g(t)$ to the end of the recording. 

We then define the absolute origin of time $t=0$ as the last maximum of $g(t)$ before impact such that the end time of the recording is given by
$$t_f = \Delta t + n T_F, \quad n \in \mathbb{Z} $$
in this reference time (figure~\ref{fig:synchro}). The impact time $t_i \in [0, T_F]$ is then calculated as
\begin{equation}
t_i = \Delta t + n T_F - (t_f - t_i) = (t_i - t_f + \Delta t)[T_F]
\end{equation}
where $X[T_F]$ means $X$ modulo $T_F$. 

The impact phase $\tau_i$ is then
\begin{equation}
 \tau_i = \frac{2 \pi t_i}{T_F} =  \frac{\left(\Delta t + (t_i-t_f)\right)\,[T_F]}{T_F}
 \label{eq:impact_phase}
\end{equation}
Practically, $t_f-t_i$ is computed as a number of frames divided by the acquisition frequency. 

The digital output of the DACQ is refreshed at 8~kHz, so the error on $\Delta t$ is less than 0.125~ms. The impact is detected with an accuracy of ($\pm 1$ frames), which corresponds to an error on $t_f - t_i$ less than 0.25~ms. The absolute error on $t_i$ (resp. $\tau_i$) is then less than 0.375~ms (resp. 1.5\%). 

\bibliographystyle{jfm}
\bibliography{bibliography}

\end{document}